\documentclass[aps,pra,reprint,superscriptaddress,showpacs]{revtex4-1}
\usepackage{graphicx}
\usepackage{amsmath}
\usepackage{textcomp}
\usepackage{color}
\usepackage{xcolor}
\usepackage{upgreek}
\usepackage{comment}
\usepackage{verbatim}
\usepackage{amssymb}
\usepackage{tabu}
\usepackage{float}
\usepackage[english]{babel}
\usepackage{gensymb}
\usepackage{natbib}
\begin{document}
\setlength{\abovedisplayskip}{5pt}
\setlength{\belowdisplayskip}{5pt}


\title{Huygens' Dipole for Polarization-Controlled Nanoscale Light Routing }



\author{Sergey Nechayev}
\affiliation{Max Planck Institute for the Science of Light, Staudtstr. 2, D-91058 Erlangen, Germany}
\affiliation{Institute of Optics, Information and Photonics, University Erlangen-Nuremberg, Staudtstr. 7/B2, D-91058 Erlangen, Germany}
\author{J\"org S. Eismann}
\affiliation{Max Planck Institute for the Science of Light, Staudtstr. 2, D-91058 Erlangen, Germany}
\affiliation{Institute of Optics, Information and Photonics, University Erlangen-Nuremberg, Staudtstr. 7/B2, D-91058 Erlangen, Germany}
\author{Martin Neugebauer}
\affiliation{Max Planck Institute for the Science of Light, Staudtstr. 2, D-91058 Erlangen, Germany}
\affiliation{Institute of Optics, Information and Photonics, University Erlangen-Nuremberg, Staudtstr. 7/B2, D-91058 Erlangen, Germany}
\author{Pawe{\l} Wo{\'z}niak}
\affiliation{Max Planck Institute for the Science of Light, Staudtstr. 2, D-91058 Erlangen, Germany}
\affiliation{Institute of Optics, Information and Photonics, University Erlangen-Nuremberg, Staudtstr. 7/B2, D-91058 Erlangen, Germany}
\author{Ankan Bag}
\affiliation{Max Planck Institute for the Science of Light, Staudtstr. 2, D-91058 Erlangen, Germany}
\affiliation{Institute of Optics, Information and Photonics, University Erlangen-Nuremberg, Staudtstr. 7/B2, D-91058 Erlangen, Germany}
\author{Gerd Leuchs}
\affiliation{Max Planck Institute for the Science of Light, Staudtstr. 2, D-91058 Erlangen, Germany}
\affiliation{Institute of Optics, Information and Photonics, University Erlangen-Nuremberg, Staudtstr. 7/B2, D-91058 Erlangen, Germany}
\author{Peter Banzer}
\email[]{peter.banzer@mpl.mpg.de}
\homepage[]{http://www.mpl.mpg.de/}
\affiliation{Max Planck Institute for the Science of Light, Staudtstr. 2, D-91058 Erlangen, Germany}
\affiliation{Institute of Optics, Information and Photonics, University Erlangen-Nuremberg, Staudtstr. 7/B2, D-91058 Erlangen, Germany}


\date{\today}

\begin{abstract}
Structured illumination allows for satisfying the first Kerker condition of in-phase perpendicular electric and magnetic dipole moments in any isotropic scatterer that supports electric and magnetic dipolar resonances. The induced Huygens' dipole may be utilized for unidirectional coupling to waveguide modes that propagate transverse to the excitation beam. We study two configurations of a Huygens' dipole -- longitudinal electric and transverse magnetic dipole moments or vice versa. We experimentally show that only the radially polarized emission of the first and azimuthally polarized emission of the second configuration are directional in the far-field. This polarization selectivity implies that directional excitation of either TM or TE waveguide modes is possible. Applying this concept to a single nanoantenna excited with structured light, we are able to experimentally achieve scattering directivities of around 23 dB and 18 dB in TM and TE modes, respectively. This strong directivity paves the way for tunable polarization-controlled nanoscale light routing and applications in optical metrology, localization microscopy and on-chip optical devices.
\end{abstract}

\pacs{03.50.De, 42.25.Ja, 42.50.Tx}

\maketitle

\section{Introduction}
In 1983, Kerker et al. predicted directional antenna-like scattering by magneto-dielectric particles~\cite{kerker_electromagnetic_1983} owing to the interference of the excited in-phase perpendicular electric and magnetic dipole moments. Such dipole source was termed Huygens' dipole and it is associated with increased forward$\backslash$suppressed backward scattering~\cite{geffrin_magnetic_2012,person_demonstration_2013}. The prediction of directional scattering under plane-wave illumination, supported by experimental demonstrations~\cite{geffrin_magnetic_2012, fu_directional_2013,person_demonstration_2013}, and multipolar generalized versions of it~\cite{alaee_generalized_2015, pors_unidirectional_2015, wei_adding_2017,liu_generalized_2017, kruk_functional_2017} turned out to be seminal in the field of optical antennas~\cite{muhlschlegel_resonant_2005, bharadwaj_optical_2009, novotny_antennas_2011,krasnok_all-dielectric_2012,krasnok_optical_2013}. 
Considerable efforts have been made to confine the radiated power by single nano-antennas~\cite{nieto-vesperinas_angle-suppressed_2011,coenen_directional_2011,rolly_boosting_2012, coenen_directional_2014,wiecha_strongly_2017,picardi_janus_2018} and metasurfaces~\cite{staude_tailoring_2013,decker_manuel_highefficiency_2015,arslan_angle-selective_2017,langguth_plasmonic_2015} into an even narrower angular range. These directive nano-antennas coupled to single emitters allow for controlling the emission intensity distribution~\cite{gersen_influencing_2000, taminiau_optical_2008, taminiau_enhanced_2008,curto_unidirectional_2010, curto_multipolar_2013,hancu_multipolar_2014,langguth_plasmonic_2015} and polarization~\cite{kruk_spin-polarized_2014, ren_linearly_2015,cotrufo_spin-dependent_2016,yan_twisting_2017}.\\
Inducing a Huygens' dipole in a nanoparticle under plane-wave illumination requires that the nanoparticle has equal first order electric $a_1$ and magnetic $b_1$ Mie coefficients~\cite{kerker_electromagnetic_1983}. Importantly, a dipolar scatterer responds only to the \textit{local} electric and magnetic fields, while Maxwell's equations locally permit \textit{any} configuration of these electromagnetic field vectors~\cite{Abouraddy_Three-Dimensional_2006,yang_role}. Consequently, structured illumination allows for exciting~\cite{wozniak_selective_2015} an arbitrary oriented Huygens' dipole in \textit{any} isotropic dipolar scatterer (assuming $a_1,\,b_1\neq 0$) and, hence, unidirectional scattering along \textit{any} axis. The first experimental observation of a Huygens' dipole that emits light directionally transverse to the propagation direction of the excitation beam was reported recently~\cite{neugebauer_polarization-controlled_2016, ankan2018} using structured illumination and a Si nanoparticle with $|a_1| \neq |b_1|$, while the relative phase between Mie coefficients $Arg (b_1/a_1) \approx \pi/2$ compensated for the inherent phase of the structured excitation field~\cite{neugebauer_polarization_2014,aiello_transverse_2015}. The observed phenomena was referred to as \textquotedblleft transverse Kerker scattering\textquotedblright.\\
In this communication, we report on an experimental polarization resolved quantitative study of transverse Kerker scattering phenomena using a spatially varying cylindrical polarization basis. We utilize two possible realizations of a Huygens' dipole --- an in-phase longitudinal electric and transverse magnetic dipole moments (\textit{i}) and vice versa (\textit{ii}) --- excited in a high-refractive index dielectric spherical nanoparticle ($a_1 \neq b_1$) with structured illumination~\cite{neugebauer_polarization-controlled_2016, ankan2018}. Employing a spatially varying cylindrical polarization basis, we directly confirm that transversely directional light emission of (\textit{i}) and (\textit{ii}) appears in the radially and azimuthally polarized (TM and TE) components~\cite{picardi_janus_2018}, respectively. We obtain transverse scattering asymmetries of approximately 23 dB and 18 dB in radially and azimuthally polarized components of the scattered light and conclude that (\textit{i}) and (\textit{ii}) are capable of directional excitation of TM and TE waveguide modes, respectively. Experimentally achieving these remarkable transverse scattering asymmetries in the specific polarization modes using an individual nanoantenna provides for a route towards tunable polarization-controlled nanoscale light routing for applications in optical metrology and on-chip optical devices.\\
\section{Theory}
We start by briefly describing the excitation of a Huygens' dipole in a Mie scatterer with structured illumination. Consider a radially polarized beam, which is tightly focused by an aplanatic objective with high numerical aperture (NA)~\cite{dorn_sharper_2003}. In the vicinity of its geometrical focus the field can be approximated by $\mathbf{E}^\text{foc}_\text{rad}\left(x,y,z\right) \propto\left(x\mathbf{e}_{x}+y\mathbf{e}_{y} +  \frac{2\imath}{k_{\text{eff}}} \mathbf{e}_{z}\right) \exp \left( \imath k_{\text{eff}}z \right)$~\cite{novotny_principles_2012, weak2}, with the Cartesian basis vectors $\mathbf{e}_\zeta$ ($\zeta=x,y,z$) and $k_{\text{eff}}$ the effective wavenumber~\cite{k_note}. We employ the Maxwell-Faraday equation for time harmonic electromagentic waves, $\mathbf{H}^\text{foc}_\text{rad}=\frac{-\imath}{k\eta}\nabla\times\mathbf{E}^\text{foc}_\text{rad}$, to obtain the focal magnetic field, where $k$ and $\eta$ are the free-space wavenumber and impedance, respectively. Finally, we obtain the focal fields of a focused azimuthally polarized beam via the electromagnetic duality transformation $\left\{\mathbf{E,H}\right\} \rightarrow \left\{\mathbf{H,-E}\right\}$. This approach leads to compact approximations of both beams in the focal plane ($z=0$) as follows:
\begin{align}\label{eqn:focal_fields1}
&\mathbf{E}^\text{foc}_\text{rad}\left(x,y\right) =  x\mathbf{e}_{x}+y\mathbf{e}_{y} +  \frac{2\imath}{k_{\text{eff}}} \mathbf{e}_{z} 
\text{,}\\ \label{eqn:focal_fields2}
&\mathbf{H}^\text{foc}_\text{rad}\left(x,y\right) = \frac{k_\text{eff}}{k \eta}\left(-y\mathbf{e}_{x}+x\mathbf{e}_{y} \right) 
\text{,}\\ \label{eqn:focal_fields3}
&\mathbf{E}^\text{foc}_\text{azi}\left(x,y\right) = y\mathbf{e}_{x}-x\mathbf{e}_{y}
\text{,}\\ \label{eqn:focal_fields4}
&\mathbf{H}^\text{foc}_\text{azi}\left(x,y\right) =  \frac{k_\text{eff}}{k \eta}\left( x\mathbf{e}_{x}+y\mathbf{e}_{y} +  \frac{2\imath}{k_{\text{eff}}} \mathbf{e}_{z} \right)
\text{.}
\end{align}
We notice in Eqs.~\eqref{eqn:focal_fields1}-\eqref{eqn:focal_fields4} that for the radially (azimuthally) polarized beam the longitudinal electric (magnetic) field is $\pm \pi/2$ dephased relatively to transverse magnetic (electric) field~\cite{neugebauer_polarization_2014,aiello_transverse_2015}. Consequently, as it was discussed in depth in~\cite{neugebauer_polarization-controlled_2016,ankan2018}, a dipolar Mie scatterer excited at a wavelength such that the Mie coefficients compensate this phase $Arg(b_1/a_1)=\pi/2$ and positioned in the proximity of the optical axis (owing to the cylindrical symmetry we only discuss positions along the $x$-axis) allows for constructing the Huygens' dipoles (\textit{i}) and (\textit{ii}) using a focused radially and azimuthally polarized beam, respectively~\cite{neugebauer_polarization-controlled_2016,ankan2018,dipoles_note}. Specifically, a radially polarized beam [Eqs.~\eqref{eqn:focal_fields1}-\eqref{eqn:focal_fields2}] excites \mbox{ $\mathbf{p} = -\imath x |a_1| \mathbf{e}_{x} + \frac{2}{k_{\text{eff}}} |a_1| \mathbf{e}_{z} \equiv \left( p_x^{\text{rad}},0,p_z^{\text{rad}} \right)$ }and \mbox{$\mathbf{m}/c = \frac{ k_{\text{eff}}}{k} x|b_1| \mathbf{e}_{y} \equiv \left( 0,m_y^{\text{rad}} /c,0\right) $}~\cite{neugebauer_polarization_2014,neugebauer_polarization-controlled_2016,weak1}, where $c$ is the speed of light in vacuum. Here, $p_z^{\text{rad}}$ and $m_y^{\text{rad}}$ constitute a Huygens' dipole (\textit{i}) at the position $x_i = \pm \frac{2|a_1|k}{|b_1|k^2_\text{eff}}$ while $p_x^{\text{rad}}$ is an unwanted parasitical component, whose contribution is minimized if $  \left| a_1 \left( \lambda \right)/b_1 \left( \lambda \right)\right|  \ll 1$~\cite{neugebauer_polarization-controlled_2016,fields_note2}. Alternatively, an azimuthally polarized beam [Eqs.~\eqref{eqn:focal_fields3}-\eqref{eqn:focal_fields4}] excites \mbox{ $\mathbf{p} = x |a_1|  \mathbf{e}_{y} \equiv \left( 0,p_y^{\text{azi}} ,0\right)$} and \mbox{$\mathbf{m}/c = -\imath \frac{k_{\text{eff}}}{k} x|b_1|\mathbf{e}_{x}+   \frac{2 }{k} |b_1|\mathbf{e}_{z} \equiv \left( m_x^{\text{azi}} /c,0,m_z^{\text{azi}} /c \right) $}~\cite{ankan2018}. In this case, $m_z^{\text{azi}}$ and $p_y^{\text{azi}}$ constitute a Huygens' dipole (\textit{ii}) at $x_{ii}=\pm \frac{2|b_1|}{|a_1|k}$, while $m_x^{\text{azi}}$ is the unwanted component, whose contribution is minimized if $  \left| a_1 \left( \lambda \right)/b_1 \left( \lambda \right)\right|  \gg 1$~\cite{ankan2018,fields_note2}.\\
To mimic the coupling of the Huygens' dipoles (\textit{i}) and (\textit{ii}) to a high-refractive index dielectric waveguide, we assume that the focal plane ($z=0$) constitutes a boundary between two media -- air ($z<0$) and dielectric ($z>0$) with a refractive index $n$. The excited dipoles are positioned in air at distance $d$ above the interface ($z<0$ half-space). The far-field scattered light $\mathbf{E}^\infty \left(k_{x},k_{y}\right) = \left[ \mathrm{E}^{\text{TM}}, \mathrm{E}^{\text{TE}} \right]^\text{T}$, coupled to the higher-density medium ($z>0$) in TM$\backslash$TE polarization basis can be written as~\cite{novotny_principles_2012,martin_magnetic}:
\begin{align}
\label{eqn:far_fields1}
&\left[\begin{matrix}
\mathrm{E}^{\text{TM}}_{(\textit{i})}\\
\mathrm{E}^{\text{TE}}_{(\textit{i})}\\
\end{matrix}\right]
\propto C \mathbf{F}
\left[\begin{matrix}
 \frac{k_xk_z}{k_\bot k} p_x^{\text{rad}} -\frac{k_\bot}{k} p_z^{\text{rad}}+ \frac{k_x}{k_\bot} m_y^{\text{rad}} /c\\
-\frac{k_{y}k_z}{k_{\bot}k}m_y^{\text{rad}} /c- \frac{k_y}{k_\bot} p_x^{\text{rad}} \\
\end{matrix}\right]
\text{,}\\
\label{eqn:far_fields2}
&\left[\begin{matrix}
\mathrm{E}^{\text{TM}}_{(\textit{ii})}\\
\mathrm{E}^{\text{TE}}_{(\textit{ii})}\\
\end{matrix}\right]
\propto C \mathbf{F}
\left[\begin{matrix}
\frac{k_{y}k_z}{k_{\bot}k}p_y^{\text{azi}} - \frac{k_{y}}{k_{\bot}}m_x^{\text{azi}}  /c \\
\frac{k_x}{k_\bot} p_y^{\text{azi}}-\frac{k_xk_z}{k_\bot k} m_x^{\text{azi}}/c +\frac{k_\bot}{k} m_z^{\text{azi}} /c \\
\end{matrix}\right]
\text{,}
\end{align}
where $\mathbf{F}= \text{diag}(t_p,t_s)$ is the matrix of the Fresnel transmission coefficients~\cite{novotny_principles_2012}, $C=\exp \left( \imath k_z d \right) \left(k^{2}n^{2}-k_{\bot}^{2}\right)^{1/2}/k_z$, \mbox{$k_{\bot}=\left(k_{x}^2+k_{y}^2\right)^{1/2}\leq nk$} is the transverse wavenumber and $k_{z}=\left(k^{2}-k_{\bot}^{2}\right)^{1/2}$ has a positive imaginary part $\Im\left[k_{z}\right]\geq0$. Eqs.~\eqref{eqn:far_fields1}-\eqref{eqn:far_fields2} show that when neglecting the small transverse dipole components $p_x^{\text{rad}}$ and $m_x^{\text{azi}}$, (\textit{i}) and (\textit{ii}) can have directionality only in TM and TE polarized emission, respectively.\\
\section{Experiment}
\begin{figure}[!tb]
  \includegraphics[width=0.48\textwidth]{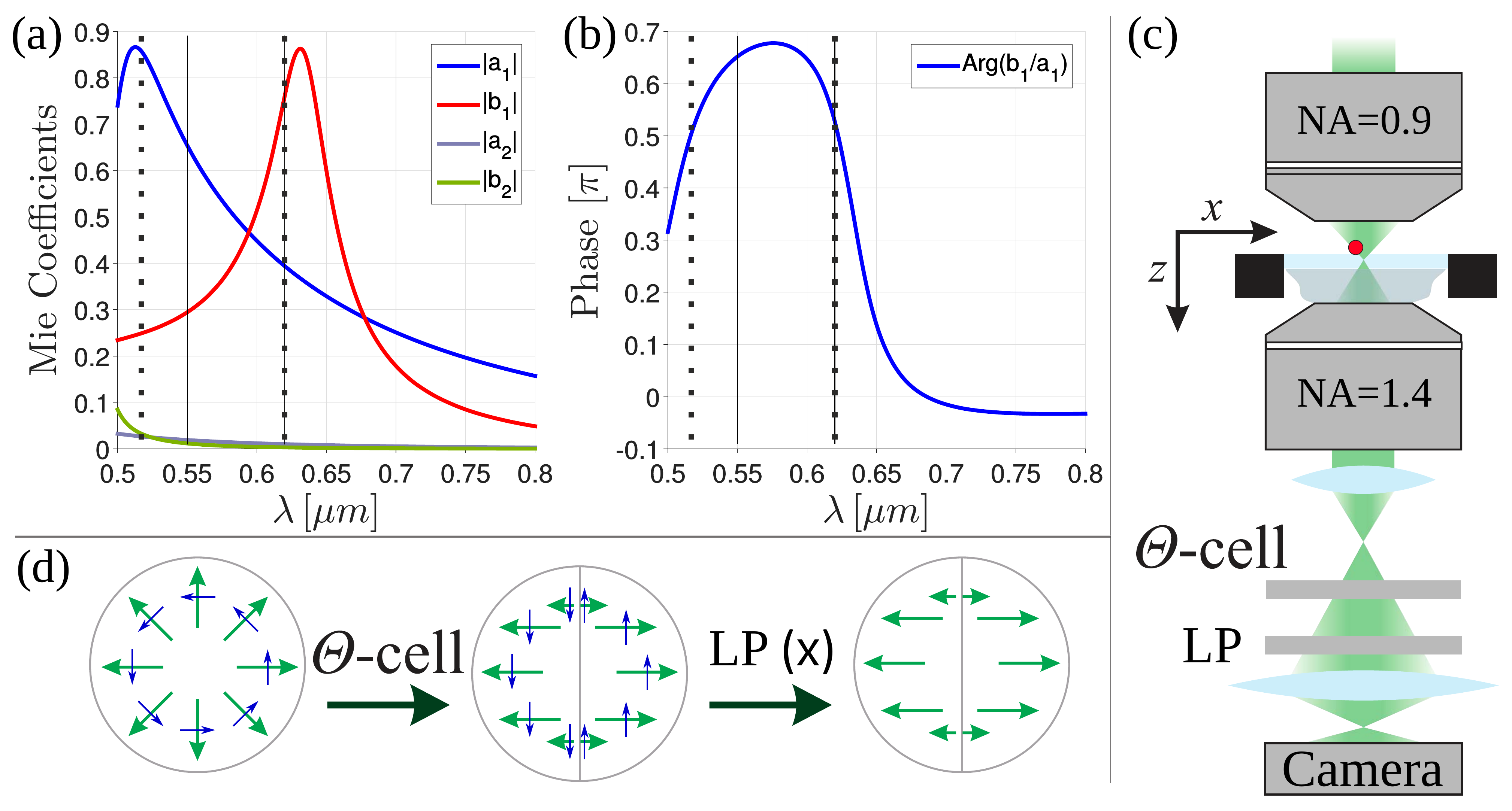}
  \caption{(a) Absolute values of the first and second order Mie coefficients of a core-shell nanoparticle with Si core of radius $r_{\text{Si}} = 78$\,nm and a SiO$_2$ shell of thickness $6$\,nm in free-space. (b) The corresponding phase difference between the first order Mie coefficients. The dotted vertical black lines show the wavelengths $\lambda_i=620 \, \mathrm{nm}$ and $\lambda_{ii}=520 \, \mathrm{nm}$ where the phase difference is approximately $\pi/2$. The solid vertical black lines show the experimentally found wavelengths $\lambda_i^{\text{exp}}=620 \, \mathrm{nm}$ and $\lambda_{ii}^{\text{exp}}=550 \, \mathrm{nm}$ that give maximum transverse scattering asymmetry into TM and TE polarized modes, respectively. (c) Experimental setup. An incident radially or azimuthally polarized beam is focused by a high numerical aperture (NA) objective onto the nanoparticle, positioned on a glass substrate, which is actuated by a 3D piezo stage. An immersion-type objective collects the transmitted and scattered light. The back focal plane of the collecting objective is imaged onto the polarization conversion-projection unit and imaged again onto a camera. (d) Schematic presentation of the polarization conversion-projection unit. The $\Uptheta$-cell converts the impinging TM (radial) and TE (azimuthal) polarizations into linearly polarized Cartesian omponents ($x$ and $y$), respectively, introducing a $\pi$-phase singularity along $x=0$. The intensity distribution in the TM and TE modes can be measured by setting the transmission axis of the subsequent linear polarizer (LP) to $x$ and $y$, respectively, as shown for $x$.}
  \label{fig:fig1}
\end{figure}

\begin{figure}[!htb]
  \includegraphics[width=0.48\textwidth]{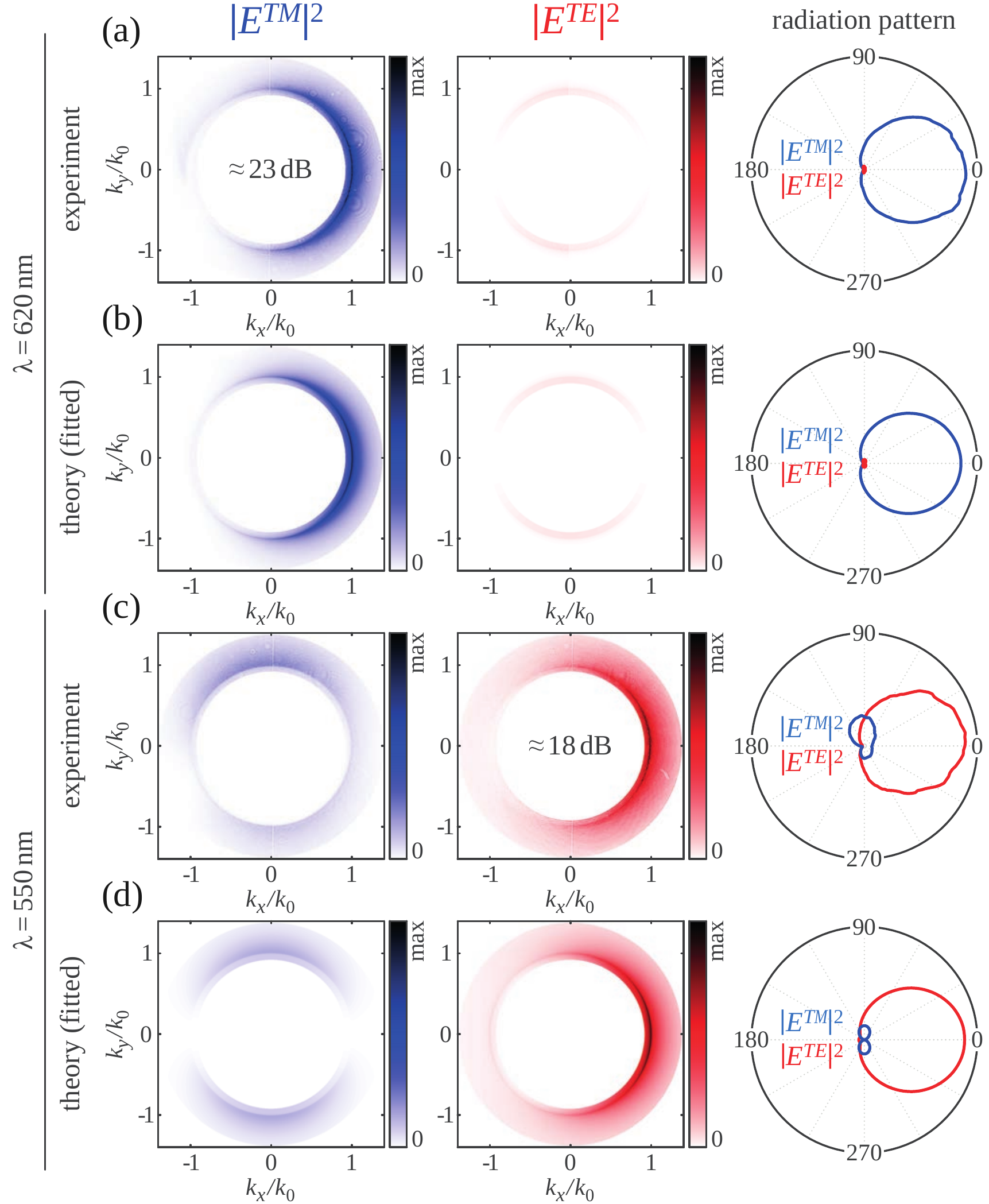}
  \caption{(a) and (c) -- polarization resolved measurements of the far-field scattered light. The nanoantenna is positioned in the focal plane of focused radially and azimuthally polarized beams at $\mathbf{r}=(-150\,\mathrm{nm},0,0)$ and $\mathbf{r}=(75\,\mathrm{nm},0,0)$, respectively. The plots show the intensity distribution in TM (radial) and TE (azimuthal) polarized modes in the back focal plane (BFP) of the collecting objective. (b) and (d) -- corresponding theoretical (fitted) BFP distributions calculated with Eqs.~\eqref{eqn:far_fields1}-\eqref{eqn:far_fields2}. The dipole moments used to plot (b) and (d)  were obtained by a nonlinear least squares fitting of (a) and (c) with Eqs.~\eqref{eqn:far_fields1}-\eqref{eqn:far_fields2}. The right column shows the corresponding radiation diagrams -- the intensity distribution in the BFP as a function of the polar angle $\theta = \tan^{-1}(k_y/k_x)$.} 
  \label{fig:experiment}
\end{figure}

As nanoantenna we choose a concentric core-shell spherical nanoparticle, positioned on a glass substrate ($n= 1.52$) using a custom AFM-based pick-and-place method~\cite{mick_pick}, with the core radius of $r_\mathrm{Si}= 78\, \mathrm{nm} $ made of crystalline silicon~\cite{Palik1985} and a $6\,\mathrm{nm}$ thick shell made of $\mathrm{SiO_2}$~\cite{Palik1985}. In Fig.~\ref{fig:fig1}~(a), we plot the first and second order free-space Mie coefficients~\cite{huffman_absorption_1983,Hightower1988} $a_1,\,b_1,\,a_2,\,b_2$, showing that the scatterer is well characterized by its dipolar response for $\lambda>520\,\mathrm{nm}$. The plot in Fig.~\ref{fig:fig1}~(b) shows that $Arg(b_1/a_1)=\pi/2$ at $\lambda_i=620 \, \mathrm{nm}$ and $\lambda_{ii}=520 \, \mathrm{nm}$, while $|a_1(\lambda_{i})/b_1(\lambda_{i})|<1$ and $|a_1(\lambda_{ii})/b_1(\lambda_{ii})|>1$ in Fig.~\ref{fig:fig1}~(a). Consequently, following our arguments in the previous section and the detailed discussion in~\cite{neugebauer_polarization-controlled_2016,ankan2018}, $\lambda_i$ and $\lambda_{ii}$ satisfy the conditions for the excitation of the Huygens' dipoles (\textit{i}) and (\textit{ii}) using focused radially and azimuthally polarized beams, respectively, while the phase delay between the Mie coefficients $Arg(b_1/a_1) = \pi/2$ is compensated by the phase delay of the exciting field components as appears in Eq.~\eqref{eqn:focal_fields1}-~\eqref{eqn:focal_fields4}.\\
Our setup~\cite{banzer_experimental_2010,neugebauer_polarization-controlled_2016} is schematically depicted in Fig.~\ref{fig:fig1}~(c). We prepare radially and azimuthally polarized beams using a q-plate~\cite{marrucci_optical_2006} and spatially filter them~\cite{karimi_hypergeometric-gaussian_2007}. The resulting beams are focused by a high numerical aperture ($\mathrm{NA}=0.9$) objective. The substrate is mounted onto a 3D piezo actuator, allowing for precise positioning of the nanoparticle in the focal plane. The transmitted and scattered light is collected by a confocally aligned index-matched immersion-type objective ($\mathrm{NA}=1.4$). The Fourier space (far-field) of the light emitted by the nanoparticle is obtained in the back focal plane (BFP) of the collecting objective up to the transverse wavenumber $k_{\bot}/k\leq 1.4$. The angular range $0.9 < k_{\bot}/k\leq 1.4$ corresponds to scattered light only. We image the BFP of the collecting objective onto a polarization conversion-projection unit consisting of a $\Uptheta$-cell~\cite{stalder_linearly_1996} and a rotatable linear polarizer. The $\Uptheta$-cell converts the TM and TE polarized components into linear Cartesian $x$ and $y$ polarized components, respectively, as illustrated in Fig.~\ref{fig:fig1}~(d). The subsequent linear polarizer filters the desired projection. The second lens in Fig.~\ref{fig:fig1}~(c) images the BFP and resulting intensity distribution onto a camera.\\
Experimentally, we record $|\mathrm{E}^{\text{TM}}|^2$ and $|\mathrm{E}^{\text{TE}}|^2$ for radially and azimuthally polarized excitations at various wavelengths and transverse positions of the nanoparticle.  We find a maximum transverse scattering asymmetry in the TM mode with radially polarized excitation at a wavelength of $\lambda_i^{\text{exp}}= 620 \, \mathrm{nm}$ at the position $|x_i^{\text{exp}}|= 150\, \mathrm{nm} \pm 5\, \mathrm{nm}$ with expected theoretical values $\lambda_i= 620 \,\mathrm{nm}$ and $|x_i|= 160 \,\mathrm{nm}$. For azimuthally polarized excitation the maximal asymmetry in the TE mode is found at $\lambda_{ii}^{\text{exp}}= 550 \, \mathrm{nm}$ and $|x_{ii}^{\text{exp}}|= 75\, \mathrm{nm} \pm 5\, \mathrm{nm}$ with theoretical values $\lambda_{ii}= 520 \, \mathrm{nm}$ and $|x_{ii}|= 78 \,\mathrm{nm}$. The deviation from the theoretical values for radially polarized excitation originates from neglecting the reflected incident field in Eqs.~\eqref{eqn:focal_fields1}-\eqref{eqn:focal_fields2}, neglecting the transverse electric field in Eq.~\eqref{eqn:focal_fields1}, substrate-induced bi-anisotropy~\cite{substrate_halas,substrate_bianisotropy} and linear approximation of the focal fields. For azimuthally polarized excitation, even larger deviations are expected owing to the rising quadrupole contributions at shorter wavelengths as seen in Fig.~\ref{fig:fig1}~(a). The mentioned effects can be incorporated in the model using a T-matrix approach~\cite{mish, ankan2018}.\\
In Fig.~\ref{fig:experiment} we plot the obtained polarization resolved BFP images at $\lambda_i^{\text{exp}}= 620 \,\mathrm{nm} $, $x_i^{\text{exp}}=-150 \, \mathrm{nm}$ for radially polarized excitation (a) and at $\lambda_{ii}^{\text{exp}}= 550  \,\mathrm{nm}$, $x_{ii}^{\text{exp}}= 75\, \mathrm{nm}$  for azimuthally polarized excitation (c). Fig.~\ref{fig:experiment} (a), (c) clearly show that the Huygens' dipole configurations (\textit{i}) and (\textit{ii}) lead to directional scattering in the TM component only for radial excitation (\textit{i}) and the TE component only (\textit{ii}) for azimuthal excitation. Hence, the two configurations are also capable of directionally exciting TM and TE polarized waveguide modes, respectively. The actual transverse scattering asymmetry into a specific polarization mode can be presented by plotting the radiation diagrams. To this end, we evaluate the TE and TM intensity components in the BFP images at the critical angle $k_\bot = k$ using polar plots \mbox{$\rho \left( \theta \right) = |\mathrm{E}^{\mathrm{TE \backslash TM}}|^2 \left( \theta \right)$}, as shown on the right-hand side of Fig.~\ref{fig:experiment}, where $\rho$ is the absolute value of radius vector and \mbox{$ \theta = \tan^{-1} (k_y/k_x)$}. We experimentally obtain directivity of approximately 23 dB and 18 dB of coupling of the Huygens' dipoles (\textit{i}) and (\textit{ii}) into TM and TE polarized modes, respectively.\\
Finally, we perform a nonlinear least squares fitting of the experimental BFP images in Fig.~\ref{fig:experiment} (a) and (c) with our model in Eqs.~\eqref{eqn:far_fields1} and \eqref{eqn:far_fields2} and summarize the results in Table~\ref{tab:_tab1}. For the wavelength of $\lambda_i^{\text{exp}}= 620 \,\mathrm{nm} $, we achieve a virtually perfect Huygens' dipole (\textit{i}) --- $p_z^{rad}  \approx -m_y^{rad}/c$ with a very significant parasitic component of $p_x^{rad}$. Nevertheless, $p_x^{rad}$ does not influence transverse scattering asymmetry into the TM polarized mode, since around the critical angle $k_\bot \approx k$ the longitudinal wavenumber $k_z \approx 0$ nullifies the contribution of $p_x^{rad}$, as appears in the first line of Eq.~(\ref{eqn:far_fields1}). Moreover, $p_x^{rad}$ also does not influence the transverse scattering asymmetry into the cross-polarized TE mode, since along the scattering direction defined by the $k_x$-axis we have $k_y \equiv 0$, which nullifies the contribution of $p_x^{rad}$ to the TE mode, as appears in the second line of Eq.~(\ref{eqn:far_fields1}). For the wavelength of $\lambda_{ii}^{\text{exp}}= 550 \,\mathrm{nm} $, we achieve a virtually perfect Huygens' dipole (\textit{ii}) --- $p_y^{azi}  \approx m_z^{azi}/c$ with a significant parasitic component of $m_x^{azi}/c$, which, following the same line of arguments as for $p_x^{rad}$, does not influence the transverse scattering asymmetry. In Fig.~\ref{fig:experiment} (b) and (d) we plot the corresponding theoretical (fitted) BFP images and radiation patterns, using the results shown in Tab.~\ref{tab:_tab1} and Eqs.~\eqref{eqn:far_fields1}-\eqref{eqn:far_fields2}.\\

\begin{table}[!h]
\centering
\def\arraystretch{1.5}
\begin{tabular}{|c|c|c|c|c|c|c|} 
\hline
$\lambda\,[\mathrm{nm}]$ &  $p_x$ &$p_y$ &$p_z$ & $m_x/c$& $m_y/c$& $m_z/c$ \\

\hline

620&$0.60 e^{ \imath 0.51 \pi}$ &0 & 1 & 0 &$0.95 e^{\imath 0.93 \pi}$& 0\\

\hline
550&0&$1.05e^{ \imath 1.84 \pi}$&0& $ 0.44 e^{\imath 1.25 \pi}$ &0&1\\

\hline

	\end{tabular}
 \caption{Summary of the dipole moments retrieved from the experimental BFP data shown in Fig.~\ref{fig:experiment} (a), (c) and fitted with Eqs.~\eqref{eqn:far_fields1}-\eqref{eqn:far_fields2}.}
 \label{tab:_tab1}
\end{table}

%
%
%
%

\section{Conclusion}
In conclusion, we have experimentally investigated the polarization properties of Huygens' dipoles induced by structured illumination by analyzing their emission properties in cylindrical polarization basis. Utilizing a single nanoantenna excited with structured light, we were able to experimentally achieve transverse scattering asymmetries of around 23 dB and 18 dB in the radial (TM) and azimuthal (TE) polarization mode, respectively. Our scheme may find applications in optical metrology, localization microscopy and on-chip tunable polarization-controlled light routing.


\begin{thebibliography}{0}%
\makeatletter
\providecommand \@ifxundefined [1]{%
 \@ifx{#1\undefined}
}%
\providecommand \@ifnum [1]{%
 \ifnum #1\expandafter \@firstoftwo
 \else \expandafter \@secondoftwo
 \fi
}%
\providecommand \@ifx [1]{%
 \ifx #1\expandafter \@firstoftwo
 \else \expandafter \@secondoftwo
 \fi
}%
\providecommand \natexlab [1]{#1}%
\providecommand \enquote  [1]{``#1''}%
\providecommand \bibnamefont  [1]{#1}%
\providecommand \bibfnamefont [1]{#1}%
\providecommand \citenamefont [1]{#1}%
\providecommand \href@noop [0]{\@secondoftwo}%
\providecommand \href [0]{\begingroup \@sanitize@url \@href}%
\providecommand \@href[1]{\@@startlink{#1}\@@href}%
\providecommand \@@href[1]{\endgroup#1\@@endlink}%
\providecommand \@sanitize@url [0]{\catcode `\\12\catcode `\$12\catcode
  `\&12\catcode `\#12\catcode `\^12\catcode `\_12\catcode `\%12\relax}%
\providecommand \@@startlink[1]{}%
\providecommand \@@endlink[0]{}%
\providecommand \url  [0]{\begingroup\@sanitize@url \@url }%
\providecommand \@url [1]{\endgroup\@href {#1}{\urlprefix }}%
\providecommand \urlprefix  [0]{URL }%
\providecommand \Eprint [0]{\href }%
\providecommand \doibase [0]{http://dx.doi.org/}%
\providecommand \selectlanguage [0]{\@gobble}%
\providecommand \bibinfo  [0]{\@secondoftwo}%
\providecommand \bibfield  [0]{\@secondoftwo}%
\providecommand \translation [1]{[#1]}%
\providecommand \BibitemOpen [0]{}%
\providecommand \bibitemStop [0]{}%
\providecommand \bibitemNoStop [0]{.\EOS\space}%
\providecommand \EOS [0]{\spacefactor3000\relax}%
\providecommand \BibitemShut  [1]{\csname bibitem#1\endcsname}%
\let\auto@bib@innerbib\@empty
\bibitem [{\citenamefont {Kerker}\ \emph {et~al.}(1983)\citenamefont {Kerker},
  \citenamefont {Wang},\ and\ \citenamefont
  {Giles}}]{kerker_electromagnetic_1983}%
  \BibitemOpen
  \bibfield  {author} {\bibinfo {author} {\bibfnamefont {M.}~\bibnamefont
  {Kerker}}, \bibinfo {author} {\bibfnamefont {D.-S.}\ \bibnamefont {Wang}}, \
  and\ \bibinfo {author} {\bibfnamefont {C.~L.}\ \bibnamefont {Giles}},\ }\href
  {\doibase 10.1364/JOSA.73.000765} {\bibfield  {journal} {\bibinfo  {journal}
  {JOSA}\ }\textbf {\bibinfo {volume} {73}},\ \bibinfo {pages} {765} (\bibinfo
  {year} {1983})}\BibitemShut {NoStop}%
\bibitem [{\citenamefont {Geffrin}\ \emph {et~al.}(2012)\citenamefont
  {Geffrin}, \citenamefont {Garc\'ia-C\'amara}, \citenamefont {G\'omez-Medina},
  \citenamefont {Albella}, \citenamefont {Froufe-P\'erez}, \citenamefont
  {Eyraud}, \citenamefont {Litman}, \citenamefont {Vaillon}, \citenamefont
  {Gonz\'alez}, \citenamefont {Nieto-Vesperinas}, \citenamefont {S\'aenz},\
  and\ \citenamefont {Moreno}}]{geffrin_magnetic_2012}%
  \BibitemOpen
  \bibfield  {author} {\bibinfo {author} {\bibfnamefont {J.~M.}\ \bibnamefont
  {Geffrin}}, \bibinfo {author} {\bibfnamefont {B.}~\bibnamefont
  {Garc\'ia-C\'amara}}, \bibinfo {author} {\bibfnamefont {R.}~\bibnamefont
  {G\'omez-Medina}}, \bibinfo {author} {\bibfnamefont {P.}~\bibnamefont
  {Albella}}, \bibinfo {author} {\bibfnamefont {L.~S.}\ \bibnamefont
  {Froufe-P\'erez}}, \bibinfo {author} {\bibfnamefont {C.}~\bibnamefont
  {Eyraud}}, \bibinfo {author} {\bibfnamefont {A.}~\bibnamefont {Litman}},
  \bibinfo {author} {\bibfnamefont {R.}~\bibnamefont {Vaillon}}, \bibinfo
  {author} {\bibfnamefont {F.}~\bibnamefont {Gonz\'alez}}, \bibinfo {author}
  {\bibfnamefont {M.}~\bibnamefont {Nieto-Vesperinas}}, \bibinfo {author}
  {\bibfnamefont {J.~J.}\ \bibnamefont {S\'aenz}}, \ and\ \bibinfo {author}
  {\bibfnamefont {F.}~\bibnamefont {Moreno}},\ }\href {\doibase
  10.1038/ncomms2167} {\bibfield  {journal} {\bibinfo  {journal} {Nature
  Communications}\ }\textbf {\bibinfo {volume} {3}},\ \bibinfo {pages} {1171}
  (\bibinfo {year} {2012})}\BibitemShut {NoStop}%
\bibitem [{\citenamefont {Person}\ \emph {et~al.}(2013)\citenamefont {Person},
  \citenamefont {Jain}, \citenamefont {Lapin}, \citenamefont {S\'aenz},
  \citenamefont {Wicks},\ and\ \citenamefont
  {Novotny}}]{person_demonstration_2013}%
  \BibitemOpen
  \bibfield  {author} {\bibinfo {author} {\bibfnamefont {S.}~\bibnamefont
  {Person}}, \bibinfo {author} {\bibfnamefont {M.}~\bibnamefont {Jain}},
  \bibinfo {author} {\bibfnamefont {Z.}~\bibnamefont {Lapin}}, \bibinfo
  {author} {\bibfnamefont {J.~J.}\ \bibnamefont {S\'aenz}}, \bibinfo {author}
  {\bibfnamefont {G.}~\bibnamefont {Wicks}}, \ and\ \bibinfo {author}
  {\bibfnamefont {L.}~\bibnamefont {Novotny}},\ }\href {\doibase
  10.1021/nl4005018} {\bibfield  {journal} {\bibinfo  {journal} {Nano Letters}\
  }\textbf {\bibinfo {volume} {13}},\ \bibinfo {pages} {1806} (\bibinfo {year}
  {2013})}\BibitemShut {NoStop}%
\bibitem [{\citenamefont {Fu}\ \emph {et~al.}(2013)\citenamefont {Fu},
  \citenamefont {Kuznetsov}, \citenamefont {Miroshnichenko}, \citenamefont
  {Yu},\ and\ \citenamefont {Luk{'}yanchuk}}]{fu_directional_2013}%
  \BibitemOpen
  \bibfield  {author} {\bibinfo {author} {\bibfnamefont {Y.~H.}\ \bibnamefont
  {Fu}}, \bibinfo {author} {\bibfnamefont {A.~I.}\ \bibnamefont {Kuznetsov}},
  \bibinfo {author} {\bibfnamefont {A.~E.}\ \bibnamefont {Miroshnichenko}},
  \bibinfo {author} {\bibfnamefont {Y.~F.}\ \bibnamefont {Yu}}, \ and\ \bibinfo
  {author} {\bibfnamefont {B.}~\bibnamefont {Luk{'}yanchuk}},\ }\href {\doibase
  10.1038/ncomms2538} {\bibfield  {journal} {\bibinfo  {journal} {Nature
  Communications}\ }\textbf {\bibinfo {volume} {4}},\ \bibinfo {pages} {1527}
  (\bibinfo {year} {2013})}\BibitemShut {NoStop}%
\bibitem [{\citenamefont {Alaee}\ \emph {et~al.}(2015)\citenamefont {Alaee},
  \citenamefont {Filter}, \citenamefont {Lehr}, \citenamefont {Lederer},\ and\
  \citenamefont {Rockstuhl}}]{alaee_generalized_2015}%
  \BibitemOpen
  \bibfield  {author} {\bibinfo {author} {\bibfnamefont {R.}~\bibnamefont
  {Alaee}}, \bibinfo {author} {\bibfnamefont {R.}~\bibnamefont {Filter}},
  \bibinfo {author} {\bibfnamefont {D.}~\bibnamefont {Lehr}}, \bibinfo {author}
  {\bibfnamefont {F.}~\bibnamefont {Lederer}}, \ and\ \bibinfo {author}
  {\bibfnamefont {C.}~\bibnamefont {Rockstuhl}},\ }\href {\doibase
  10.1364/OL.40.002645} {\bibfield  {journal} {\bibinfo  {journal} {Optics
  Letters}\ }\textbf {\bibinfo {volume} {40}},\ \bibinfo {pages} {2645}
  (\bibinfo {year} {2015})}\BibitemShut {NoStop}%
\bibitem [{\citenamefont {Pors}\ \emph {et~al.}(2015)\citenamefont {Pors},
  \citenamefont {Andersen},\ and\ \citenamefont
  {Bozhevolnyi}}]{pors_unidirectional_2015}%
  \BibitemOpen
  \bibfield  {author} {\bibinfo {author} {\bibfnamefont {A.}~\bibnamefont
  {Pors}}, \bibinfo {author} {\bibfnamefont {S.~K.~H.}\ \bibnamefont
  {Andersen}}, \ and\ \bibinfo {author} {\bibfnamefont {S.~I.}\ \bibnamefont
  {Bozhevolnyi}},\ }\href {\doibase 10.1364/OE.23.028808} {\bibfield  {journal}
  {\bibinfo  {journal} {Optics Express}\ }\textbf {\bibinfo {volume} {23}},\
  \bibinfo {pages} {28808} (\bibinfo {year} {2015})}\BibitemShut {NoStop}%
\bibitem [{\citenamefont {Wei}\ \emph {et~al.}(2017)\citenamefont {Wei},
  \citenamefont {Bhattacharya},\ and\ \citenamefont
  {Urbach}}]{wei_adding_2017}%
  \BibitemOpen
  \bibfield  {author} {\bibinfo {author} {\bibfnamefont {L.}~\bibnamefont
  {Wei}}, \bibinfo {author} {\bibfnamefont {N.}~\bibnamefont {Bhattacharya}}, \
  and\ \bibinfo {author} {\bibfnamefont {H.~P.}\ \bibnamefont {Urbach}},\
  }\href {\doibase 10.1364/OL.42.001776} {\bibfield  {journal} {\bibinfo
  {journal} {Optics Letters}\ }\textbf {\bibinfo {volume} {42}},\ \bibinfo
  {pages} {1776} (\bibinfo {year} {2017})}\BibitemShut {NoStop}%
\bibitem [{\citenamefont {Liu}\ and\ \citenamefont
  {Kivshar}(2018)}]{liu_generalized_2017}%
  \BibitemOpen
  \bibfield  {author} {\bibinfo {author} {\bibfnamefont {W.}~\bibnamefont
  {Liu}}\ and\ \bibinfo {author} {\bibfnamefont {Y.~S.}\ \bibnamefont
  {Kivshar}},\ }\href {\doibase 10.1364/OE.26.013085} {\bibfield  {journal}
  {\bibinfo  {journal} {Opt. Express}\ }\textbf {\bibinfo {volume} {26}},\
  \bibinfo {pages} {13085} (\bibinfo {year} {2018})}\BibitemShut {NoStop}%
\bibitem [{\citenamefont {Kruk}\ and\ \citenamefont
  {Kivshar}(2017)}]{kruk_functional_2017}%
  \BibitemOpen
  \bibfield  {author} {\bibinfo {author} {\bibfnamefont {S.}~\bibnamefont
  {Kruk}}\ and\ \bibinfo {author} {\bibfnamefont {Y.}~\bibnamefont {Kivshar}},\
  }\href {\doibase 10.1021/acsphotonics.7b01038} {\bibfield  {journal}
  {\bibinfo  {journal} {ACS Photonics}\ }\textbf {\bibinfo {volume} {4}},\
  \bibinfo {pages} {2638} (\bibinfo {year} {2017})}\BibitemShut {NoStop}%
\bibitem [{\citenamefont {M{\"{u}}hlschlegel}\ \emph
  {et~al.}(2005)\citenamefont {M{\"{u}}hlschlegel}, \citenamefont {Eisler},
  \citenamefont {Martin}, \citenamefont {Hecht},\ and\ \citenamefont
  {Pohl}}]{muhlschlegel_resonant_2005}%
  \BibitemOpen
  \bibfield  {author} {\bibinfo {author} {\bibfnamefont {P.}~\bibnamefont
  {M{\"{u}}hlschlegel}}, \bibinfo {author} {\bibfnamefont {H.-J.}\ \bibnamefont
  {Eisler}}, \bibinfo {author} {\bibfnamefont {O.~J.~F.}\ \bibnamefont
  {Martin}}, \bibinfo {author} {\bibfnamefont {B.}~\bibnamefont {Hecht}}, \
  and\ \bibinfo {author} {\bibfnamefont {D.~W.}\ \bibnamefont {Pohl}},\ }\href
  {\doibase 10.1126/science.1111886} {\bibfield  {journal} {\bibinfo  {journal}
  {Science}\ }\textbf {\bibinfo {volume} {308}},\ \bibinfo {pages} {1607}
  (\bibinfo {year} {2005})}\BibitemShut {NoStop}%
\bibitem [{\citenamefont {Bharadwaj}\ \emph {et~al.}(2009)\citenamefont
  {Bharadwaj}, \citenamefont {Deutsch},\ and\ \citenamefont
  {Novotny}}]{bharadwaj_optical_2009}%
  \BibitemOpen
  \bibfield  {author} {\bibinfo {author} {\bibfnamefont {P.}~\bibnamefont
  {Bharadwaj}}, \bibinfo {author} {\bibfnamefont {B.}~\bibnamefont {Deutsch}},
  \ and\ \bibinfo {author} {\bibfnamefont {L.}~\bibnamefont {Novotny}},\ }\href
  {\doibase 10.1364/AOP.1.000438} {\bibfield  {journal} {\bibinfo  {journal}
  {Advances in Optics and Photonics}\ }\textbf {\bibinfo {volume} {1}},\
  \bibinfo {pages} {438} (\bibinfo {year} {2009})}\BibitemShut {NoStop}%
\bibitem [{\citenamefont {Novotny}\ and\ \citenamefont {van
  Hulst}(2011)}]{novotny_antennas_2011}%
  \BibitemOpen
  \bibfield  {author} {\bibinfo {author} {\bibfnamefont {L.}~\bibnamefont
  {Novotny}}\ and\ \bibinfo {author} {\bibfnamefont {N.}~\bibnamefont {van
  Hulst}},\ }\href {\doibase 10.1038/nphoton.2010.237} {\bibfield  {journal}
  {\bibinfo  {journal} {Nature Photonics}\ }\textbf {\bibinfo {volume} {5}},\
  \bibinfo {pages} {83} (\bibinfo {year} {2011})}\BibitemShut {NoStop}%
\bibitem [{\citenamefont {Krasnok}\ \emph {et~al.}(2012)\citenamefont
  {Krasnok}, \citenamefont {Miroshnichenko}, \citenamefont {Belov},\ and\
  \citenamefont {Kivshar}}]{krasnok_all-dielectric_2012}%
  \BibitemOpen
  \bibfield  {author} {\bibinfo {author} {\bibfnamefont {A.~E.}\ \bibnamefont
  {Krasnok}}, \bibinfo {author} {\bibfnamefont {A.~E.}\ \bibnamefont
  {Miroshnichenko}}, \bibinfo {author} {\bibfnamefont {P.~A.}\ \bibnamefont
  {Belov}}, \ and\ \bibinfo {author} {\bibfnamefont {Y.~S.}\ \bibnamefont
  {Kivshar}},\ }\href {\doibase 10.1364/OE.20.020599} {\bibfield  {journal}
  {\bibinfo  {journal} {Optics Express}\ }\textbf {\bibinfo {volume} {20}},\
  \bibinfo {pages} {20599} (\bibinfo {year} {2012})}\BibitemShut {NoStop}%
\bibitem [{\citenamefont {Krasnok}\ \emph {et~al.}(2013)\citenamefont
  {Krasnok}, \citenamefont {Maksymov}, \citenamefont {Denisyuk}, \citenamefont
  {Belov}, \citenamefont {Miroshnichenko}, \citenamefont {Simovski},\ and\
  \citenamefont {Kivshar}}]{krasnok_optical_2013}%
  \BibitemOpen
  \bibfield  {author} {\bibinfo {author} {\bibfnamefont {A.~E.}\ \bibnamefont
  {Krasnok}}, \bibinfo {author} {\bibfnamefont {I.~S.}\ \bibnamefont
  {Maksymov}}, \bibinfo {author} {\bibfnamefont {A.~I.}\ \bibnamefont
  {Denisyuk}}, \bibinfo {author} {\bibfnamefont {P.~A.}\ \bibnamefont {Belov}},
  \bibinfo {author} {\bibfnamefont {A.~E.}\ \bibnamefont {Miroshnichenko}},
  \bibinfo {author} {\bibfnamefont {C.~R.}\ \bibnamefont {Simovski}}, \ and\
  \bibinfo {author} {\bibfnamefont {Y.~S.}\ \bibnamefont {Kivshar}},\ }\href
  {\doibase 10.3367/UFNe.0183.201306a.0561} {\bibfield  {journal} {\bibinfo
  {journal} {Physics-Uspekhi}\ }\textbf {\bibinfo {volume} {56}},\ \bibinfo
  {pages} {539} (\bibinfo {year} {2013})}\BibitemShut {NoStop}%
\bibitem [{\citenamefont {Nieto-Vesperinas}\ \emph {et~al.}(2011)\citenamefont
  {Nieto-Vesperinas}, \citenamefont {Gomez-Medina},\ and\ \citenamefont
  {Saenz}}]{nieto-vesperinas_angle-suppressed_2011}%
  \BibitemOpen
  \bibfield  {author} {\bibinfo {author} {\bibfnamefont {M.}~\bibnamefont
  {Nieto-Vesperinas}}, \bibinfo {author} {\bibfnamefont {R.}~\bibnamefont
  {Gomez-Medina}}, \ and\ \bibinfo {author} {\bibfnamefont {J.~J.}\
  \bibnamefont {Saenz}},\ }\href {\doibase 10.1364/JOSAA.28.000054} {\bibfield
  {journal} {\bibinfo  {journal} {JOSA A}\ }\textbf {\bibinfo {volume} {28}},\
  \bibinfo {pages} {54} (\bibinfo {year} {2011})}\BibitemShut {NoStop}%
\bibitem [{\citenamefont {Coenen}\ \emph {et~al.}(2011)\citenamefont {Coenen},
  \citenamefont {Vesseur}, \citenamefont {Polman},\ and\ \citenamefont
  {Koenderink}}]{coenen_directional_2011}%
  \BibitemOpen
  \bibfield  {author} {\bibinfo {author} {\bibfnamefont {T.}~\bibnamefont
  {Coenen}}, \bibinfo {author} {\bibfnamefont {E.~J.~R.}\ \bibnamefont
  {Vesseur}}, \bibinfo {author} {\bibfnamefont {A.}~\bibnamefont {Polman}}, \
  and\ \bibinfo {author} {\bibfnamefont {A.~F.}\ \bibnamefont {Koenderink}},\
  }\href {\doibase 10.1021/nl201839g} {\bibfield  {journal} {\bibinfo
  {journal} {Nano Letters}\ }\textbf {\bibinfo {volume} {11}},\ \bibinfo
  {pages} {3779} (\bibinfo {year} {2011})}\BibitemShut {NoStop}%
\bibitem [{\citenamefont {Rolly}\ \emph {et~al.}(2012)\citenamefont {Rolly},
  \citenamefont {Stout},\ and\ \citenamefont {Bonod}}]{rolly_boosting_2012}%
  \BibitemOpen
  \bibfield  {author} {\bibinfo {author} {\bibfnamefont {B.}~\bibnamefont
  {Rolly}}, \bibinfo {author} {\bibfnamefont {B.}~\bibnamefont {Stout}}, \ and\
  \bibinfo {author} {\bibfnamefont {N.}~\bibnamefont {Bonod}},\ }\href
  {\doibase 10.1364/OE.20.020376} {\bibfield  {journal} {\bibinfo  {journal}
  {Optics Express}\ }\textbf {\bibinfo {volume} {20}},\ \bibinfo {pages}
  {20376} (\bibinfo {year} {2012})}\BibitemShut {NoStop}%
\bibitem [{\citenamefont {Coenen}\ \emph {et~al.}(2014)\citenamefont {Coenen},
  \citenamefont {Bernal~Arango}, \citenamefont {Femius~Koenderink},\ and\
  \citenamefont {Polman}}]{coenen_directional_2014}%
  \BibitemOpen
  \bibfield  {author} {\bibinfo {author} {\bibfnamefont {T.}~\bibnamefont
  {Coenen}}, \bibinfo {author} {\bibfnamefont {F.}~\bibnamefont
  {Bernal~Arango}}, \bibinfo {author} {\bibfnamefont {A.}~\bibnamefont
  {Femius~Koenderink}}, \ and\ \bibinfo {author} {\bibfnamefont
  {A.}~\bibnamefont {Polman}},\ }\href {\doibase 10.1038/ncomms4250} {\bibfield
   {journal} {\bibinfo  {journal} {Nature Communications}\ }\textbf {\bibinfo
  {volume} {5}},\ \bibinfo {pages} {3250} (\bibinfo {year} {2014})}\BibitemShut
  {NoStop}%
\bibitem [{\citenamefont {Wiecha}\ \emph {et~al.}(2017)\citenamefont {Wiecha},
  \citenamefont {Cuche}, \citenamefont {Arbouet}, \citenamefont {Girard},
  \citenamefont {Colas~des Francs}, \citenamefont {Lecestre}, \citenamefont
  {Larrieu}, \citenamefont {Fournel}, \citenamefont {Larrey}, \citenamefont
  {Baron},\ and\ \citenamefont {Paillard}}]{wiecha_strongly_2017}%
  \BibitemOpen
  \bibfield  {author} {\bibinfo {author} {\bibfnamefont {P.~R.}\ \bibnamefont
  {Wiecha}}, \bibinfo {author} {\bibfnamefont {A.}~\bibnamefont {Cuche}},
  \bibinfo {author} {\bibfnamefont {A.}~\bibnamefont {Arbouet}}, \bibinfo
  {author} {\bibfnamefont {C.}~\bibnamefont {Girard}}, \bibinfo {author}
  {\bibfnamefont {G.}~\bibnamefont {Colas~des Francs}}, \bibinfo {author}
  {\bibfnamefont {A.}~\bibnamefont {Lecestre}}, \bibinfo {author}
  {\bibfnamefont {G.}~\bibnamefont {Larrieu}}, \bibinfo {author} {\bibfnamefont
  {F.}~\bibnamefont {Fournel}}, \bibinfo {author} {\bibfnamefont
  {V.}~\bibnamefont {Larrey}}, \bibinfo {author} {\bibfnamefont
  {T.}~\bibnamefont {Baron}}, \ and\ \bibinfo {author} {\bibfnamefont
  {V.}~\bibnamefont {Paillard}},\ }\href {\doibase
  10.1021/acsphotonics.7b00423} {\bibfield  {journal} {\bibinfo  {journal} {ACS
  Photonics}\ }\textbf {\bibinfo {volume} {4}},\ \bibinfo {pages} {2036}
  (\bibinfo {year} {2017})}\BibitemShut {NoStop}%
\bibitem [{\citenamefont {Picardi}\ \emph {et~al.}(2018)\citenamefont
  {Picardi}, \citenamefont {Zayats},\ and\ \citenamefont
  {Rodr{\'{i}}guez-Fortu{\~{n}}o}}]{picardi_janus_2018}%
  \BibitemOpen
  \bibfield  {author} {\bibinfo {author} {\bibfnamefont {M.~F.}\ \bibnamefont
  {Picardi}}, \bibinfo {author} {\bibfnamefont {A.~V.}\ \bibnamefont {Zayats}},
  \ and\ \bibinfo {author} {\bibfnamefont {F.~J.}\ \bibnamefont
  {Rodr{\'{i}}guez-Fortu{\~{n}}o}},\ }\href {\doibase
  10.1103/PhysRevLett.120.117402} {\bibfield  {journal} {\bibinfo  {journal}
  {Physical Review Letters}\ }\textbf {\bibinfo {volume} {120}},\ \bibinfo
  {pages} {117402} (\bibinfo {year} {2018})}\BibitemShut {NoStop}%
\bibitem [{\citenamefont {Staude}\ \emph {et~al.}(2013)\citenamefont {Staude},
  \citenamefont {Miroshnichenko}, \citenamefont {Decker}, \citenamefont
  {Fofang}, \citenamefont {Liu}, \citenamefont {Gonzales}, \citenamefont
  {Dominguez}, \citenamefont {Luk}, \citenamefont {Neshev}, \citenamefont
  {Brener},\ and\ \citenamefont {Kivshar}}]{staude_tailoring_2013}%
  \BibitemOpen
  \bibfield  {author} {\bibinfo {author} {\bibfnamefont {I.}~\bibnamefont
  {Staude}}, \bibinfo {author} {\bibfnamefont {A.~E.}\ \bibnamefont
  {Miroshnichenko}}, \bibinfo {author} {\bibfnamefont {M.}~\bibnamefont
  {Decker}}, \bibinfo {author} {\bibfnamefont {N.~T.}\ \bibnamefont {Fofang}},
  \bibinfo {author} {\bibfnamefont {S.}~\bibnamefont {Liu}}, \bibinfo {author}
  {\bibfnamefont {E.}~\bibnamefont {Gonzales}}, \bibinfo {author}
  {\bibfnamefont {J.}~\bibnamefont {Dominguez}}, \bibinfo {author}
  {\bibfnamefont {T.~S.}\ \bibnamefont {Luk}}, \bibinfo {author} {\bibfnamefont
  {D.~N.}\ \bibnamefont {Neshev}}, \bibinfo {author} {\bibfnamefont
  {I.}~\bibnamefont {Brener}}, \ and\ \bibinfo {author} {\bibfnamefont
  {Y.}~\bibnamefont {Kivshar}},\ }\href {\doibase 10.1021/nn402736f} {\bibfield
   {journal} {\bibinfo  {journal} {ACS Nano}\ }\textbf {\bibinfo {volume}
  {7}},\ \bibinfo {pages} {7824} (\bibinfo {year} {2013})}\BibitemShut
  {NoStop}%
\bibitem [{\citenamefont {{Decker Manuel}}\ \emph {et~al.}(2015)\citenamefont
  {{Decker Manuel}}, \citenamefont {{Staude Isabelle}}, \citenamefont {{Falkner
  Matthias}}, \citenamefont {{Dominguez Jason}}, \citenamefont {{Neshev
  Dragomir N.}}, \citenamefont {{Brener Igal}}, \citenamefont {{Pertsch
  Thomas}},\ and\ \citenamefont {{Kivshar Yuri
  S.}}}]{decker_manuel_highefficiency_2015}%
  \BibitemOpen
  \bibfield  {author} {\bibinfo {author} {\bibnamefont {{Decker Manuel}}},
  \bibinfo {author} {\bibnamefont {{Staude Isabelle}}}, \bibinfo {author}
  {\bibnamefont {{Falkner Matthias}}}, \bibinfo {author} {\bibnamefont
  {{Dominguez Jason}}}, \bibinfo {author} {\bibnamefont {{Neshev Dragomir
  N.}}}, \bibinfo {author} {\bibnamefont {{Brener Igal}}}, \bibinfo {author}
  {\bibnamefont {{Pertsch Thomas}}}, \ and\ \bibinfo {author} {\bibnamefont
  {{Kivshar Yuri S.}}},\ }\href {\doibase 10.1002/adom.201400584} {\bibfield
  {journal} {\bibinfo  {journal} {Advanced Optical Materials}\ }\textbf
  {\bibinfo {volume} {3}},\ \bibinfo {pages} {813} (\bibinfo {year}
  {2015})}\BibitemShut {NoStop}%
\bibitem [{\citenamefont {Arslan}\ \emph {et~al.}(2017)\citenamefont {Arslan},
  \citenamefont {Chong}, \citenamefont {Miroshnichenko}, \citenamefont {Choi},
  \citenamefont {Neshev}, \citenamefont {Pertsch}, \citenamefont {Kivshar},\
  and\ \citenamefont {Staude}}]{arslan_angle-selective_2017}%
  \BibitemOpen
  \bibfield  {author} {\bibinfo {author} {\bibfnamefont {D.}~\bibnamefont
  {Arslan}}, \bibinfo {author} {\bibfnamefont {K.~E.}\ \bibnamefont {Chong}},
  \bibinfo {author} {\bibfnamefont {A.~E.}\ \bibnamefont {Miroshnichenko}},
  \bibinfo {author} {\bibfnamefont {D.-Y.}\ \bibnamefont {Choi}}, \bibinfo
  {author} {\bibfnamefont {D.~N.}\ \bibnamefont {Neshev}}, \bibinfo {author}
  {\bibfnamefont {T.}~\bibnamefont {Pertsch}}, \bibinfo {author} {\bibfnamefont
  {Y.~S.}\ \bibnamefont {Kivshar}}, \ and\ \bibinfo {author} {\bibfnamefont
  {I.}~\bibnamefont {Staude}},\ }\href {\doibase 10.1088/1361-6463/aa875c}
  {\bibfield  {journal} {\bibinfo  {journal} {Journal of Physics D: Applied
  Physics}\ }\textbf {\bibinfo {volume} {50}},\ \bibinfo {pages} {434002}
  (\bibinfo {year} {2017})}\BibitemShut {NoStop}%
\bibitem [{\citenamefont {Langguth}\ \emph {et~al.}(2015)\citenamefont
  {Langguth}, \citenamefont {Schokker}, \citenamefont {Guo},\ and\
  \citenamefont {Koenderink}}]{langguth_plasmonic_2015}%
  \BibitemOpen
  \bibfield  {author} {\bibinfo {author} {\bibfnamefont {L.}~\bibnamefont
  {Langguth}}, \bibinfo {author} {\bibfnamefont {A.~H.}\ \bibnamefont
  {Schokker}}, \bibinfo {author} {\bibfnamefont {K.}~\bibnamefont {Guo}}, \
  and\ \bibinfo {author} {\bibfnamefont {A.~F.}\ \bibnamefont {Koenderink}},\
  }\href {\doibase 10.1103/PhysRevB.92.205401} {\bibfield  {journal} {\bibinfo
  {journal} {Physical Review B}\ }\textbf {\bibinfo {volume} {92}},\ \bibinfo
  {pages} {205401} (\bibinfo {year} {2015})}\BibitemShut {NoStop}%
\bibitem [{\citenamefont {Gersen}\ \emph {et~al.}(2000)\citenamefont {Gersen},
  \citenamefont {Garc\'ia-Paraj\'o}, \citenamefont {Novotny}, \citenamefont
  {Veerman}, \citenamefont {Kuipers},\ and\ \citenamefont {van
  Hulst}}]{gersen_influencing_2000}%
  \BibitemOpen
  \bibfield  {author} {\bibinfo {author} {\bibfnamefont {H.}~\bibnamefont
  {Gersen}}, \bibinfo {author} {\bibfnamefont {M.~F.}\ \bibnamefont
  {Garc\'ia-Paraj\'o}}, \bibinfo {author} {\bibfnamefont {L.}~\bibnamefont
  {Novotny}}, \bibinfo {author} {\bibfnamefont {J.~A.}\ \bibnamefont
  {Veerman}}, \bibinfo {author} {\bibfnamefont {L.}~\bibnamefont {Kuipers}}, \
  and\ \bibinfo {author} {\bibfnamefont {N.~F.}\ \bibnamefont {van Hulst}},\
  }\href {\doibase 10.1103/PhysRevLett.85.5312} {\bibfield  {journal} {\bibinfo
   {journal} {Physical Review Letters}\ }\textbf {\bibinfo {volume} {85}},\
  \bibinfo {pages} {5312} (\bibinfo {year} {2000})}\BibitemShut {NoStop}%
\bibitem [{\citenamefont {Taminiau}\ \emph
  {et~al.}(2008{\natexlab{a}})\citenamefont {Taminiau}, \citenamefont
  {Stefani}, \citenamefont {Segerink},\ and\ \citenamefont {van
  Hulst}}]{taminiau_optical_2008}%
  \BibitemOpen
  \bibfield  {author} {\bibinfo {author} {\bibfnamefont {T.~H.}\ \bibnamefont
  {Taminiau}}, \bibinfo {author} {\bibfnamefont {F.~D.}\ \bibnamefont
  {Stefani}}, \bibinfo {author} {\bibfnamefont {F.~B.}\ \bibnamefont
  {Segerink}}, \ and\ \bibinfo {author} {\bibfnamefont {N.~F.}\ \bibnamefont
  {van Hulst}},\ }\href {\doibase 10.1038/nphoton.2008.32} {\bibfield
  {journal} {\bibinfo  {journal} {Nature Photonics}\ }\textbf {\bibinfo
  {volume} {2}},\ \bibinfo {pages} {234} (\bibinfo {year}
  {2008}{\natexlab{a}})}\BibitemShut {NoStop}%
\bibitem [{\citenamefont {Taminiau}\ \emph
  {et~al.}(2008{\natexlab{b}})\citenamefont {Taminiau}, \citenamefont
  {Stefani},\ and\ \citenamefont {Hulst}}]{taminiau_enhanced_2008}%
  \BibitemOpen
  \bibfield  {author} {\bibinfo {author} {\bibfnamefont {T.~H.}\ \bibnamefont
  {Taminiau}}, \bibinfo {author} {\bibfnamefont {F.~D.}\ \bibnamefont
  {Stefani}}, \ and\ \bibinfo {author} {\bibfnamefont {N.~F.~v.}\ \bibnamefont
  {Hulst}},\ }\href {\doibase 10.1364/OE.16.010858} {\bibfield  {journal}
  {\bibinfo  {journal} {Optics Express}\ }\textbf {\bibinfo {volume} {16}},\
  \bibinfo {pages} {10858} (\bibinfo {year} {2008}{\natexlab{b}})}\BibitemShut
  {NoStop}%
\bibitem [{\citenamefont {Curto}\ \emph {et~al.}(2010)\citenamefont {Curto},
  \citenamefont {Volpe}, \citenamefont {Taminiau}, \citenamefont {Kreuzer},
  \citenamefont {Quidant},\ and\ \citenamefont
  {Hulst}}]{curto_unidirectional_2010}%
  \BibitemOpen
  \bibfield  {author} {\bibinfo {author} {\bibfnamefont {A.~G.}\ \bibnamefont
  {Curto}}, \bibinfo {author} {\bibfnamefont {G.}~\bibnamefont {Volpe}},
  \bibinfo {author} {\bibfnamefont {T.~H.}\ \bibnamefont {Taminiau}}, \bibinfo
  {author} {\bibfnamefont {M.~P.}\ \bibnamefont {Kreuzer}}, \bibinfo {author}
  {\bibfnamefont {R.}~\bibnamefont {Quidant}}, \ and\ \bibinfo {author}
  {\bibfnamefont {N.~F.~v.}\ \bibnamefont {Hulst}},\ }\href {\doibase
  10.1126/science.1191922} {\bibfield  {journal} {\bibinfo  {journal}
  {Science}\ }\textbf {\bibinfo {volume} {329}},\ \bibinfo {pages} {930}
  (\bibinfo {year} {2010})}\BibitemShut {NoStop}%
\bibitem [{\citenamefont {Curto}\ \emph {et~al.}(2013)\citenamefont {Curto},
  \citenamefont {Taminiau}, \citenamefont {Volpe}, \citenamefont {Kreuzer},
  \citenamefont {Quidant},\ and\ \citenamefont {van
  Hulst}}]{curto_multipolar_2013}%
  \BibitemOpen
  \bibfield  {author} {\bibinfo {author} {\bibfnamefont {A.~G.}\ \bibnamefont
  {Curto}}, \bibinfo {author} {\bibfnamefont {T.~H.}\ \bibnamefont {Taminiau}},
  \bibinfo {author} {\bibfnamefont {G.}~\bibnamefont {Volpe}}, \bibinfo
  {author} {\bibfnamefont {M.~P.}\ \bibnamefont {Kreuzer}}, \bibinfo {author}
  {\bibfnamefont {R.}~\bibnamefont {Quidant}}, \ and\ \bibinfo {author}
  {\bibfnamefont {N.~F.}\ \bibnamefont {van Hulst}},\ }\href {\doibase
  10.1038/ncomms2769} {\bibfield  {journal} {\bibinfo  {journal} {Nature
  Communications}\ }\textbf {\bibinfo {volume} {4}},\ \bibinfo {pages} {1750}
  (\bibinfo {year} {2013})}\BibitemShut {NoStop}%
\bibitem [{\citenamefont {Hancu}\ \emph {et~al.}(2014)\citenamefont {Hancu},
  \citenamefont {Curto}, \citenamefont {Castro-L\'opez}, \citenamefont
  {Kuttge},\ and\ \citenamefont {van Hulst}}]{hancu_multipolar_2014}%
  \BibitemOpen
  \bibfield  {author} {\bibinfo {author} {\bibfnamefont {I.~M.}\ \bibnamefont
  {Hancu}}, \bibinfo {author} {\bibfnamefont {A.~G.}\ \bibnamefont {Curto}},
  \bibinfo {author} {\bibfnamefont {M.}~\bibnamefont {Castro-L\'opez}},
  \bibinfo {author} {\bibfnamefont {M.}~\bibnamefont {Kuttge}}, \ and\ \bibinfo
  {author} {\bibfnamefont {N.~F.}\ \bibnamefont {van Hulst}},\ }\href {\doibase
  10.1021/nl403681g} {\bibfield  {journal} {\bibinfo  {journal} {Nano Letters}\
  }\textbf {\bibinfo {volume} {14}},\ \bibinfo {pages} {166} (\bibinfo {year}
  {2014})}\BibitemShut {NoStop}%
\bibitem [{\citenamefont {Kruk}\ \emph {et~al.}(2014)\citenamefont {Kruk},
  \citenamefont {Decker}, \citenamefont {Staude}, \citenamefont {Schlecht},
  \citenamefont {Greppmair}, \citenamefont {Neshev},\ and\ \citenamefont
  {Kivshar}}]{kruk_spin-polarized_2014}%
  \BibitemOpen
  \bibfield  {author} {\bibinfo {author} {\bibfnamefont {S.~S.}\ \bibnamefont
  {Kruk}}, \bibinfo {author} {\bibfnamefont {M.}~\bibnamefont {Decker}},
  \bibinfo {author} {\bibfnamefont {I.}~\bibnamefont {Staude}}, \bibinfo
  {author} {\bibfnamefont {S.}~\bibnamefont {Schlecht}}, \bibinfo {author}
  {\bibfnamefont {M.}~\bibnamefont {Greppmair}}, \bibinfo {author}
  {\bibfnamefont {D.~N.}\ \bibnamefont {Neshev}}, \ and\ \bibinfo {author}
  {\bibfnamefont {Y.~S.}\ \bibnamefont {Kivshar}},\ }\href {\doibase
  10.1021/ph500288u} {\bibfield  {journal} {\bibinfo  {journal} {ACS
  Photonics}\ }\textbf {\bibinfo {volume} {1}},\ \bibinfo {pages} {1218}
  (\bibinfo {year} {2014})}\BibitemShut {NoStop}%
\bibitem [{\citenamefont {Ren}\ \emph {et~al.}(2015)\citenamefont {Ren},
  \citenamefont {Chen}, \citenamefont {Wu}, \citenamefont {Zhang},
  \citenamefont {Liu}, \citenamefont {Pi}, \citenamefont {Zhang}, \citenamefont
  {Li}, \citenamefont {Fan},\ and\ \citenamefont {Xu}}]{ren_linearly_2015}%
  \BibitemOpen
  \bibfield  {author} {\bibinfo {author} {\bibfnamefont {M.}~\bibnamefont
  {Ren}}, \bibinfo {author} {\bibfnamefont {M.}~\bibnamefont {Chen}}, \bibinfo
  {author} {\bibfnamefont {W.}~\bibnamefont {Wu}}, \bibinfo {author}
  {\bibfnamefont {L.}~\bibnamefont {Zhang}}, \bibinfo {author} {\bibfnamefont
  {J.}~\bibnamefont {Liu}}, \bibinfo {author} {\bibfnamefont {B.}~\bibnamefont
  {Pi}}, \bibinfo {author} {\bibfnamefont {X.}~\bibnamefont {Zhang}}, \bibinfo
  {author} {\bibfnamefont {Q.}~\bibnamefont {Li}}, \bibinfo {author}
  {\bibfnamefont {S.}~\bibnamefont {Fan}}, \ and\ \bibinfo {author}
  {\bibfnamefont {J.}~\bibnamefont {Xu}},\ }\href {\doibase 10.1021/nl5047973}
  {\bibfield  {journal} {\bibinfo  {journal} {Nano Letters}\ }\textbf {\bibinfo
  {volume} {15}},\ \bibinfo {pages} {2951} (\bibinfo {year}
  {2015})}\BibitemShut {NoStop}%
\bibitem [{\citenamefont {Cotrufo}\ \emph {et~al.}(2016)\citenamefont
  {Cotrufo}, \citenamefont {Osorio},\ and\ \citenamefont
  {Koenderink}}]{cotrufo_spin-dependent_2016}%
  \BibitemOpen
  \bibfield  {author} {\bibinfo {author} {\bibfnamefont {M.}~\bibnamefont
  {Cotrufo}}, \bibinfo {author} {\bibfnamefont {C.~I.}\ \bibnamefont {Osorio}},
  \ and\ \bibinfo {author} {\bibfnamefont {A.~F.}\ \bibnamefont {Koenderink}},\
  }\href {\doibase 10.1021/acsnano.5b07231} {\bibfield  {journal} {\bibinfo
  {journal} {ACS Nano}\ }\textbf {\bibinfo {volume} {10}},\ \bibinfo {pages}
  {3389} (\bibinfo {year} {2016})}\BibitemShut {NoStop}%
\bibitem [{\citenamefont {Yan}\ \emph {et~al.}(2017)\citenamefont {Yan},
  \citenamefont {Wang}, \citenamefont {Raziman},\ and\ \citenamefont
  {Martin}}]{yan_twisting_2017}%
  \BibitemOpen
  \bibfield  {author} {\bibinfo {author} {\bibfnamefont {C.}~\bibnamefont
  {Yan}}, \bibinfo {author} {\bibfnamefont {X.}~\bibnamefont {Wang}}, \bibinfo
  {author} {\bibfnamefont {T.~V.}\ \bibnamefont {Raziman}}, \ and\ \bibinfo
  {author} {\bibfnamefont {O.~J.~F.}\ \bibnamefont {Martin}},\ }\href {\doibase
  10.1021/acs.nanolett.6b04906} {\bibfield  {journal} {\bibinfo  {journal}
  {Nano Letters}\ }\textbf {\bibinfo {volume} {17}},\ \bibinfo {pages} {2265}
  (\bibinfo {year} {2017})}\BibitemShut {NoStop}%
\bibitem [{\citenamefont {Abouraddy}\ and\ \citenamefont
  {Toussaint}(2006)}]{Abouraddy_Three-Dimensional_2006}%
  \BibitemOpen
  \bibfield  {author} {\bibinfo {author} {\bibfnamefont {A.~F.}\ \bibnamefont
  {Abouraddy}}\ and\ \bibinfo {author} {\bibfnamefont {K.~C.}\ \bibnamefont
  {Toussaint}},\ }\href {\doibase 10.1103/PhysRevLett.96.153901} {\bibfield
  {journal} {\bibinfo  {journal} {Phys. Rev. Lett.}\ }\textbf {\bibinfo
  {volume} {96}},\ \bibinfo {pages} {153901} (\bibinfo {year}
  {2006})}\BibitemShut {NoStop}%
\bibitem [{\citenamefont {Yang}\ and\ \citenamefont {Cohen}(2011)}]{yang_role}%
  \BibitemOpen
  \bibfield  {author} {\bibinfo {author} {\bibfnamefont {N.}~\bibnamefont
  {Yang}}\ and\ \bibinfo {author} {\bibfnamefont {A.~E.}\ \bibnamefont
  {Cohen}},\ }\href@noop {} {\bibfield  {journal} {\bibinfo  {journal} {The
  Journal of Physical Chemistry B}\ }\textbf {\bibinfo {volume} {115}},\
  \bibinfo {pages} {5304} (\bibinfo {year} {2011})}\BibitemShut {NoStop}%
\bibitem [{\citenamefont {Wo{\'z}niak}\ \emph {et~al.}(2015)\citenamefont
  {Wo{\'z}niak}, \citenamefont {Banzer},\ and\ \citenamefont
  {Leuchs}}]{wozniak_selective_2015}%
  \BibitemOpen
  \bibfield  {author} {\bibinfo {author} {\bibfnamefont {P.}~\bibnamefont
  {Wo{\'z}niak}}, \bibinfo {author} {\bibfnamefont {P.}~\bibnamefont {Banzer}},
  \ and\ \bibinfo {author} {\bibfnamefont {G.}~\bibnamefont {Leuchs}},\ }\href
  {\doibase 10.1002/lpor.201400188} {\bibfield  {journal} {\bibinfo  {journal}
  {Laser \& Photonics Reviews}\ }\textbf {\bibinfo {volume} {9}},\ \bibinfo
  {pages} {231} (\bibinfo {year} {2015})}\BibitemShut {NoStop}%
\bibitem [{\citenamefont {Neugebauer}\ \emph {et~al.}(2016)\citenamefont
  {Neugebauer}, \citenamefont {Wo{\'{z}}niak}, \citenamefont {Bag},
  \citenamefont {Leuchs},\ and\ \citenamefont
  {Banzer}}]{neugebauer_polarization-controlled_2016}%
  \BibitemOpen
  \bibfield  {author} {\bibinfo {author} {\bibfnamefont {M.}~\bibnamefont
  {Neugebauer}}, \bibinfo {author} {\bibfnamefont {P.}~\bibnamefont
  {Wo{\'{z}}niak}}, \bibinfo {author} {\bibfnamefont {A.}~\bibnamefont {Bag}},
  \bibinfo {author} {\bibfnamefont {G.}~\bibnamefont {Leuchs}}, \ and\ \bibinfo
  {author} {\bibfnamefont {P.}~\bibnamefont {Banzer}},\ }\href {\doibase
  10.1038/ncomms11286} {\bibfield  {journal} {\bibinfo  {journal} {Nature
  Communications}\ }\textbf {\bibinfo {volume} {7}},\ \bibinfo {pages} {11286}
  (\bibinfo {year} {2016})}\BibitemShut {NoStop}%
\bibitem [{\citenamefont {Bag}\ \emph {et~al.}(2018)\citenamefont {Bag},
  \citenamefont {Neugebauer}, \citenamefont {Wo\ifmmode~\acute{z}\else
  \'{z}\fi{}niak}, \citenamefont {Leuchs},\ and\ \citenamefont
  {Banzer}}]{ankan2018}%
  \BibitemOpen
  \bibfield  {author} {\bibinfo {author} {\bibfnamefont {A.}~\bibnamefont
  {Bag}}, \bibinfo {author} {\bibfnamefont {M.}~\bibnamefont {Neugebauer}},
  \bibinfo {author} {\bibfnamefont {P.}~\bibnamefont {Wo\ifmmode~\acute{z}\else
  \'{z}\fi{}niak}}, \bibinfo {author} {\bibfnamefont {G.}~\bibnamefont
  {Leuchs}}, \ and\ \bibinfo {author} {\bibfnamefont {P.}~\bibnamefont
  {Banzer}},\ }\href {\doibase 10.1103/PhysRevLett.121.193902} {\bibfield
  {journal} {\bibinfo  {journal} {Phys. Rev. Lett.}\ }\textbf {\bibinfo
  {volume} {121}},\ \bibinfo {pages} {193902} (\bibinfo {year}
  {2018})}\BibitemShut {NoStop}%
\bibitem [{\citenamefont {Neugebauer}\ \emph {et~al.}(2014)\citenamefont
  {Neugebauer}, \citenamefont {Bauer}, \citenamefont {Banzer},\ and\
  \citenamefont {Leuchs}}]{neugebauer_polarization_2014}%
  \BibitemOpen
  \bibfield  {author} {\bibinfo {author} {\bibfnamefont {M.}~\bibnamefont
  {Neugebauer}}, \bibinfo {author} {\bibfnamefont {T.}~\bibnamefont {Bauer}},
  \bibinfo {author} {\bibfnamefont {P.}~\bibnamefont {Banzer}}, \ and\ \bibinfo
  {author} {\bibfnamefont {G.}~\bibnamefont {Leuchs}},\ }\href {\doibase
  10.1021/nl5003526} {\bibfield  {journal} {\bibinfo  {journal} {Nano Letters}\
  }\textbf {\bibinfo {volume} {14}},\ \bibinfo {pages} {2546} (\bibinfo {year}
  {2014})}\BibitemShut {NoStop}%
\bibitem [{\citenamefont {Aiello}\ \emph {et~al.}(2015)\citenamefont {Aiello},
  \citenamefont {Banzer}, \citenamefont {Neugebauer},\ and\ \citenamefont
  {Leuchs}}]{aiello_transverse_2015}%
  \BibitemOpen
  \bibfield  {author} {\bibinfo {author} {\bibfnamefont {A.}~\bibnamefont
  {Aiello}}, \bibinfo {author} {\bibfnamefont {P.}~\bibnamefont {Banzer}},
  \bibinfo {author} {\bibfnamefont {M.}~\bibnamefont {Neugebauer}}, \ and\
  \bibinfo {author} {\bibfnamefont {G.}~\bibnamefont {Leuchs}},\ }\href
  {\doibase 10.1038/nphoton.2015.203} {\bibfield  {journal} {\bibinfo
  {journal} {Nature Photonics}\ }\textbf {\bibinfo {volume} {9}},\ \bibinfo
  {pages} {789} (\bibinfo {year} {2015})}\BibitemShut {NoStop}%
\bibitem [{\citenamefont {Dorn}\ \emph {et~al.}(2003)\citenamefont {Dorn},
  \citenamefont {Quabis},\ and\ \citenamefont {Leuchs}}]{dorn_sharper_2003}%
  \BibitemOpen
  \bibfield  {author} {\bibinfo {author} {\bibfnamefont {R.}~\bibnamefont
  {Dorn}}, \bibinfo {author} {\bibfnamefont {S.}~\bibnamefont {Quabis}}, \ and\
  \bibinfo {author} {\bibfnamefont {G.}~\bibnamefont {Leuchs}},\ }\href
  {\doibase 10.1103/PhysRevLett.91.233901} {\bibfield  {journal} {\bibinfo
  {journal} {Physical Review Letters}\ }\textbf {\bibinfo {volume} {91}},\
  \bibinfo {pages} {233901} (\bibinfo {year} {2003})}\BibitemShut {NoStop}%
\bibitem [{\citenamefont {Novotny}\ and\ \citenamefont
  {Hecht}(2012)}]{novotny_principles_2012}%
  \BibitemOpen
  \bibfield  {author} {\bibinfo {author} {\bibfnamefont {L.}~\bibnamefont
  {Novotny}}\ and\ \bibinfo {author} {\bibfnamefont {B.}~\bibnamefont
  {Hecht}},\ }\href@noop {} {\emph {\bibinfo {title} {Principles of
  {Nano}-{Optics}}}}\ (\bibinfo  {publisher} {Cambridge University Press},\
  \bibinfo {address} {Cambridge},\ \bibinfo {year} {2012})\BibitemShut
  {NoStop}%
\bibitem [{\citenamefont {Neugebauer}\ \emph {et~al.}(2019)\citenamefont
  {Neugebauer}, \citenamefont {Nechayev}, \citenamefont {Vorndran},
  \citenamefont {Leuchs},\ and\ \citenamefont {Banzer}}]{weak2}%
  \BibitemOpen
  \bibfield  {author} {\bibinfo {author} {\bibfnamefont {M.}~\bibnamefont
  {Neugebauer}}, \bibinfo {author} {\bibfnamefont {S.}~\bibnamefont
  {Nechayev}}, \bibinfo {author} {\bibfnamefont {M.}~\bibnamefont {Vorndran}},
  \bibinfo {author} {\bibfnamefont {G.}~\bibnamefont {Leuchs}}, \ and\ \bibinfo
  {author} {\bibfnamefont {P.}~\bibnamefont {Banzer}},\ }\href {\doibase
  10.1021/acs.nanolett.8b04219} {\bibfield  {journal} {\bibinfo  {journal}
  {Nano Letters}\ }\textbf {\bibinfo {volume} {19}},\ \bibinfo {pages} {422}
  (\bibinfo {year} {2019})}\BibitemShut {NoStop}%
\bibitem [{k_n()}]{k_note}%
  \BibitemOpen
  \href@noop {} {}\bibinfo {note} {$k_{\text{eff}}$ depends on the focusing
  system~\cite{novotny_principles_2012} and is limited by the wavenumber:
  $0\leq k_{\text{eff}}\leq k$. In paraxial approximation,
  $k_{\text{eff}}\approx k$. For the parameters of the focusing system used in
  our experiment, $k_{\text{eff}} \approx 0.74 k$ and $k_{\text{eff}} \approx
  0.78 k$ for radially and azimuthally polarized illumination,
  respectively}\BibitemShut {NoStop}%
\bibitem [{dip()}]{dipoles_note}%
  \BibitemOpen
  \href@noop {} {}\bibinfo {note} {The excited electric and magnetic dipole
  moments in the scatterer are $\mathbf{p}=\varepsilon_0 \alpha_e
  \mathbf{E}^\text{foc}\left( x \right)$ and $\mathbf{m}=
  \alpha_m\mathbf{H}^\text{foc}\left( x \right)$, where $\varepsilon_0$ is the
  vacuum permittivity. The electric ($\alpha_e$) and magnetic ($\alpha_m$)
  polarizabilites of the particle are related to the Mie coefficients as
  $\alpha_e = \left( 6 \pi \imath / k_1^3 \right)a_1$ and $\alpha_m = \left( 6
  \pi \imath / k_1^3 \right)b_1$. We omit global amplitude and phase
  factors}\BibitemShut {NoStop}%
\bibitem [{\citenamefont {Nechayev}\ \emph {et~al.}(2018)\citenamefont
  {Nechayev}, \citenamefont {Neugebauer}, \citenamefont {Vorndran},
  \citenamefont {Leuchs},\ and\ \citenamefont {Banzer}}]{weak1}%
  \BibitemOpen
  \bibfield  {author} {\bibinfo {author} {\bibfnamefont {S.}~\bibnamefont
  {Nechayev}}, \bibinfo {author} {\bibfnamefont {M.}~\bibnamefont
  {Neugebauer}}, \bibinfo {author} {\bibfnamefont {M.}~\bibnamefont
  {Vorndran}}, \bibinfo {author} {\bibfnamefont {G.}~\bibnamefont {Leuchs}}, \
  and\ \bibinfo {author} {\bibfnamefont {P.}~\bibnamefont {Banzer}},\ }\href
  {\doibase 10.1103/PhysRevLett.121.243903} {\bibfield  {journal} {\bibinfo
  {journal} {Phys. Rev. Lett.}\ }\textbf {\bibinfo {volume} {121}},\ \bibinfo
  {pages} {243903} (\bibinfo {year} {2018})}\BibitemShut {NoStop}%
\bibitem [{fie()}]{fields_note2}%
  \BibitemOpen
  \href@noop {} {}\bibinfo {note} {It should be noted that for the radially
  (azimuthally) polarized beam only the longitudinal electric (magnetic) field
  exists on the optical axis, which further fasciliates the minimisation of the
  parasitic component for $x_i \ll \lambda$ ($x_{ii} \ll \lambda$)}\BibitemShut
  {NoStop}%
\bibitem [{\citenamefont {Neugebauer}\ \emph {et~al.}(2018)\citenamefont
  {Neugebauer}, \citenamefont {Eismann}, \citenamefont {Bauer},\ and\
  \citenamefont {Banzer}}]{martin_magnetic}%
  \BibitemOpen
  \bibfield  {author} {\bibinfo {author} {\bibfnamefont {M.}~\bibnamefont
  {Neugebauer}}, \bibinfo {author} {\bibfnamefont {J.~S.}\ \bibnamefont
  {Eismann}}, \bibinfo {author} {\bibfnamefont {T.}~\bibnamefont {Bauer}}, \
  and\ \bibinfo {author} {\bibfnamefont {P.}~\bibnamefont {Banzer}},\ }\href
  {\doibase 10.1103/PhysRevX.8.021042} {\bibfield  {journal} {\bibinfo
  {journal} {Phys. Rev. X}\ }\textbf {\bibinfo {volume} {8}},\ \bibinfo {pages}
  {021042} (\bibinfo {year} {2018})}\BibitemShut {NoStop}%
\bibitem [{\citenamefont {Mick}\ \emph {et~al.}(2014)\citenamefont {Mick},
  \citenamefont {Banzer}, \citenamefont {Christiansen},\ and\ \citenamefont
  {Leuchs}}]{mick_pick}%
  \BibitemOpen
  \bibfield  {author} {\bibinfo {author} {\bibfnamefont {U.}~\bibnamefont
  {Mick}}, \bibinfo {author} {\bibfnamefont {P.}~\bibnamefont {Banzer}},
  \bibinfo {author} {\bibfnamefont {S.}~\bibnamefont {Christiansen}}, \ and\
  \bibinfo {author} {\bibfnamefont {G.}~\bibnamefont {Leuchs}},\ }in\ \href
  {\doibase 10.1364/CLEO_SI.2014.STu1H.1} {\emph {\bibinfo {booktitle} {CLEO:
  2014}}}\ (\bibinfo  {publisher} {Optical Society of America},\ \bibinfo
  {year} {2014})\ p.\ \bibinfo {pages} {STu1H.1}\BibitemShut {NoStop}%
\bibitem [{\citenamefont {Palik}(1985)}]{Palik1985}%
  \BibitemOpen
  \bibfield  {author} {\bibinfo {author} {\bibfnamefont {E.~D.}\ \bibnamefont
  {Palik}},\ }\href {https://search.library.wisc.edu/catalog/999554063402121}
  {\emph {\bibinfo {title} {{Handbook of optical constants of solids}}}}\
  (\bibinfo  {publisher} {Orlando Academic Press},\ \bibinfo {address}
  {Orlando},\ \bibinfo {year} {1985})\BibitemShut {NoStop}%
\bibitem [{\citenamefont {Bohren}\ and\ \citenamefont
  {Huffman}(1983)}]{huffman_absorption_1983}%
  \BibitemOpen
  \bibfield  {author} {\bibinfo {author} {\bibfnamefont {C.~F.}\ \bibnamefont
  {Bohren}}\ and\ \bibinfo {author} {\bibfnamefont {D.~R.}\ \bibnamefont
  {Huffman}},\ }\href@noop {} {{\selectlanguage {english}\emph {\bibinfo
  {title} {Absorption and {Scattering} of {Light} by {Small} {Particles}}}}}\
  (\bibinfo  {publisher} {John Wiley \& Sons},\ \bibinfo {address} {New York},\
  \bibinfo {year} {1983})\BibitemShut {NoStop}%
\bibitem [{\citenamefont {Hightower}\ and\ \citenamefont
  {Richardson}(1988)}]{Hightower1988}%
  \BibitemOpen
  \bibfield  {author} {\bibinfo {author} {\bibfnamefont {R.~L.}\ \bibnamefont
  {Hightower}}\ and\ \bibinfo {author} {\bibfnamefont {C.~B.}\ \bibnamefont
  {Richardson}},\ }\href {\doibase 10.1364/AO.27.004850} {\bibfield  {journal}
  {\bibinfo  {journal} {Appl. Opt.}\ }\textbf {\bibinfo {volume} {27}},\
  \bibinfo {pages} {4850} (\bibinfo {year} {1988})}\BibitemShut {NoStop}%
\bibitem [{\citenamefont {Banzer}\ \emph {et~al.}(2010)\citenamefont {Banzer},
  \citenamefont {Peschel}, \citenamefont {Quabis},\ and\ \citenamefont
  {Leuchs}}]{banzer_experimental_2010}%
  \BibitemOpen
  \bibfield  {author} {\bibinfo {author} {\bibfnamefont {P.}~\bibnamefont
  {Banzer}}, \bibinfo {author} {\bibfnamefont {U.}~\bibnamefont {Peschel}},
  \bibinfo {author} {\bibfnamefont {S.}~\bibnamefont {Quabis}}, \ and\ \bibinfo
  {author} {\bibfnamefont {G.}~\bibnamefont {Leuchs}},\ }\href {\doibase
  10.1364/OE.18.010905} {\bibfield  {journal} {\bibinfo  {journal} {Optics
  Express}\ }\textbf {\bibinfo {volume} {18}},\ \bibinfo {pages} {10905}
  (\bibinfo {year} {2010})}\BibitemShut {NoStop}%
\bibitem [{\citenamefont {Marrucci}\ \emph {et~al.}(2006)\citenamefont
  {Marrucci}, \citenamefont {Manzo},\ and\ \citenamefont
  {Paparo}}]{marrucci_optical_2006}%
  \BibitemOpen
  \bibfield  {author} {\bibinfo {author} {\bibfnamefont {L.}~\bibnamefont
  {Marrucci}}, \bibinfo {author} {\bibfnamefont {C.}~\bibnamefont {Manzo}}, \
  and\ \bibinfo {author} {\bibfnamefont {D.}~\bibnamefont {Paparo}},\ }\href
  {\doibase 10.1103/PhysRevLett.96.163905} {\bibfield  {journal} {\bibinfo
  {journal} {Physical Review Letters}\ }\textbf {\bibinfo {volume} {96}},\
  \bibinfo {pages} {163905} (\bibinfo {year} {2006})}\BibitemShut {NoStop}%
\bibitem [{\citenamefont {Karimi}\ \emph {et~al.}(2007)\citenamefont {Karimi},
  \citenamefont {Zito}, \citenamefont {Piccirillo}, \citenamefont {Marrucci},\
  and\ \citenamefont {Santamato}}]{karimi_hypergeometric-gaussian_2007}%
  \BibitemOpen
  \bibfield  {author} {\bibinfo {author} {\bibfnamefont {E.}~\bibnamefont
  {Karimi}}, \bibinfo {author} {\bibfnamefont {G.}~\bibnamefont {Zito}},
  \bibinfo {author} {\bibfnamefont {B.}~\bibnamefont {Piccirillo}}, \bibinfo
  {author} {\bibfnamefont {L.}~\bibnamefont {Marrucci}}, \ and\ \bibinfo
  {author} {\bibfnamefont {E.}~\bibnamefont {Santamato}},\ }\href {\doibase
  10.1364/OL.32.003053} {\bibfield  {journal} {\bibinfo  {journal} {Optics
  Letters}\ }\textbf {\bibinfo {volume} {32}},\ \bibinfo {pages} {3053}
  (\bibinfo {year} {2007})}\BibitemShut {NoStop}%
\bibitem [{\citenamefont {Stalder}\ and\ \citenamefont
  {Schadt}(1996)}]{stalder_linearly_1996}%
  \BibitemOpen
  \bibfield  {author} {\bibinfo {author} {\bibfnamefont {M.}~\bibnamefont
  {Stalder}}\ and\ \bibinfo {author} {\bibfnamefont {M.}~\bibnamefont
  {Schadt}},\ }\href {\doibase 10.1364/OL.21.001948} {\bibfield  {journal}
  {\bibinfo  {journal} {Optics Letters}\ }\textbf {\bibinfo {volume} {21}},\
  \bibinfo {pages} {1948} (\bibinfo {year} {1996})}\BibitemShut {NoStop}%
\bibitem [{\citenamefont {Knight}\ \emph {et~al.}(2009)\citenamefont {Knight},
  \citenamefont {Wu}, \citenamefont {Lassiter}, \citenamefont {Nordlander},\
  and\ \citenamefont {Halas}}]{substrate_halas}%
  \BibitemOpen
  \bibfield  {author} {\bibinfo {author} {\bibfnamefont {M.~W.}\ \bibnamefont
  {Knight}}, \bibinfo {author} {\bibfnamefont {Y.}~\bibnamefont {Wu}}, \bibinfo
  {author} {\bibfnamefont {J.~B.}\ \bibnamefont {Lassiter}}, \bibinfo {author}
  {\bibfnamefont {P.}~\bibnamefont {Nordlander}}, \ and\ \bibinfo {author}
  {\bibfnamefont {N.~J.}\ \bibnamefont {Halas}},\ }\href {\doibase
  10.1021/nl900945q} {\bibfield  {journal} {\bibinfo  {journal} {Nano Letters}\
  }\textbf {\bibinfo {volume} {9}},\ \bibinfo {pages} {2188} (\bibinfo {year}
  {2009})}\BibitemShut {NoStop}%
\bibitem [{\citenamefont {Miroshnichenko}\ \emph {et~al.}(2015)\citenamefont
  {Miroshnichenko}, \citenamefont {Evlyukhin}, \citenamefont {Kivshar},\ and\
  \citenamefont {Chichkov}}]{substrate_bianisotropy}%
  \BibitemOpen
  \bibfield  {author} {\bibinfo {author} {\bibfnamefont {A.~E.}\ \bibnamefont
  {Miroshnichenko}}, \bibinfo {author} {\bibfnamefont {A.~B.}\ \bibnamefont
  {Evlyukhin}}, \bibinfo {author} {\bibfnamefont {Y.~S.}\ \bibnamefont
  {Kivshar}}, \ and\ \bibinfo {author} {\bibfnamefont {B.~N.}\ \bibnamefont
  {Chichkov}},\ }\href {\doibase 10.1021/acsphotonics.5b00117} {\bibfield
  {journal} {\bibinfo  {journal} {ACS Photonics}\ }\textbf {\bibinfo {volume}
  {2}},\ \bibinfo {pages} {1423} (\bibinfo {year} {2015})}\BibitemShut
  {NoStop}%
\bibitem [{\citenamefont {Mishchenko}\ \emph {et~al.}(2002)\citenamefont
  {Mishchenko}, \citenamefont {Travis},\ and\ \citenamefont {Lacis}}]{mish}%
  \BibitemOpen
  \bibfield  {author} {\bibinfo {author} {\bibfnamefont {M.~I.}\ \bibnamefont
  {Mishchenko}}, \bibinfo {author} {\bibfnamefont {L.~D.}\ \bibnamefont
  {Travis}}, \ and\ \bibinfo {author} {\bibfnamefont {A.~A.}\ \bibnamefont
  {Lacis}},\ }\href@noop {} {\emph {\bibinfo {title} {Scattering, Absorption,
  and Emission of Light by Small Particles}}}\ (\bibinfo  {publisher}
  {Cambridge University Press},\ \bibinfo {address} {Cambridge},\ \bibinfo
  {year} {2002})\BibitemShut {NoStop}%
\end{thebibliography}

%

\end{document}